\documentclass[twocolumn]{aastex61}

\newcommand\aastex{AAS\TeX}

\received{}
\revised{}
\accepted{}
\submitjournal{ApJ}

\shorttitle{\aastex\ Grand minima and maxima from a FT-dynamo}
\shortauthors{Inceoglu et al.}
\begin{document}

\title{Nature of grand minima and maxima from fully non-linear Flux-Transport Dynamos}

\correspondingauthor{Fadil Inceoglu}
\email{finceoglu@aip.de}

\author[0000-0003-4726-3994]{Fadil Inceoglu}
\affil{Leibniz-Institute for Astrophysics Potsdam \\
An der Sternwarte 16, 14482, Potsdam, Germany}

\author{Rainer Arlt}
\affiliation{Leibniz-Institute for Astrophysics Potsdam \\
An der Sternwarte 16, 14482, Potsdam, Germany}

\author{Matthias Rempel}
\affiliation{High Altitude Observatory \\
National Center for Atmospheric Research \\
P.O. Box 3000, Boulder, CO 80307, USA}



\begin{abstract}

We aim to investigate the nature and occurrence characteristics of grand solar minimum and maximum periods, which are observed in the solar proxy records such as $^{10}$Be and $^{14}$C, using a fully non-linear Babcock-Leighton type flux-transport dynamo including momentum and entropy equations. The differential rotation and meridional circulation are generated from the effect of turbulent Reynolds stress and are subjected to back-reaction from the magnetic field. To generate grand minimum and maximum-like periods in our simulations, we used random fluctuations in the angular momentum transport process, namely the $\Lambda$-mechanism, and in the Babcock-Leighton mechanism. To characterise the nature and occurrences of the identified grand minima and maxima in our simulations, we used the waiting time distribution analyses, which reflects whether the underlying distribution arises from a random or a memory-bearing process. The results show that, in majority of the cases, the distributions of grand minima and maxima reveal that the nature of these events originates from memoryless processes. We also found that in our simulations the meridional circulation speed tends to be smaller during grand maximum, while it is faster during grand minimum periods. The radial differential rotation tend to be larger during grand maxima, while it is smaller during grand minima. The latitudinal differential rotation on the other hand is found to be larger during grand minima.

\end{abstract}




\section{Introduction} \label{sec:intro}

The Sun, the main energy source for Earth's climate, governs the space weather in the heliosphere, and shows magnetic activity structures on its photosphere, such as sunspots. Observations of sunspots since the 1610s revealed that the Sun shows a cyclic activity pattern, duration and amplitude of which vary throughout the time, the so-called Schwabe cycle. The sunspot observations also revealed that the Schwabe cycles (11-year sunspot cycles) are superimposed on a longer-term variation, where the overall activity levels of the Sun changes dramatically, such as the Maunder Minimum (1645--1715), when sunspots were almost absent on the photosphere, and the Modern Maximum (1910--2000), when the level of sunspot activity was relatively high \citep{2013LRSP...10....1U}.

Information on the solar activity levels prior to sunspot observations comes from cosmogenic radionuclide records stemming from the terrestrial archives. The most widely used cosmogenic isotopes for this purpose are $^{10}$Be in ice cores and $^{14}$C in tree rings \citep{1990Natur.347..164B,2009GeoRL..3616701K,2012PNAS..109.5967S,2015A&A...577A..20I,2016SoPh..291..303I}. The production rates of cosmogenic nuclides depend on the intensity with which cosmic rays impinge on the Earth's atmosphere \citep{Dunai2010}. However, before reaching the Earth, cosmic-ray particles have to travel through the heliosphere \citep{2013LRSP...10....3P}, where they become modulated by the open solar magnetic field \citep{2002GeoRL..29.2224L,2004SoPh..224...21W}. The production rates of the cosmogenic nuclides are inversely correlated with solar activity.

Earlier studies on the long-term variations in solar activity levels based on past production rates of cosmogenic nuclides showed that the solar activity showed quiescent and enhanced activity periods, the so-called grand solar minima and maxima, observed over the last ~10,000 years \citep{2007A&A...471..301U,2015A&A...577A..20I,2016SoPh..291..303I}. \citet{2015A&A...577A..20I} suggested that during the period from 1650 CE back to 6600 BCE, the Sun experienced 32 grand minima and 21 grand maxima, whereas \citet{2007A&A...471..301U} claimed that the Sun underwent 27 grand minima and 19 grand maxima since 8500 BCE based on their sunspot number reconstructions. Additionally, the occurrence characteristics of these periods suggest that grand solar minima and maxima are different modes in solar activity \citep{2007A&A...471..301U,2015A&A...577A..20I}. Also, \citet{2016SoPh..291..303I} suggested that $\sim$71\% of grand maxima are followed by a grand minimum during the period from 6600 BCE to 1650 CE at 93\% significance level.

The physical mechanism responsible for the generation, and the spatial and temporal evolution of the magnetic activity of the Sun is called the solar dynamo, in which small-scale flows are able to support a self-excited global magnetic field in the convective envelope of the Sun \citep{1955ApJ...122..293P,1955ApJ...121..491P}. Rotating, stratified, and electrically conducting turbulence is actually able to generate a large-scale magnetic field, in most cases by what is referred to as the  $\alpha$-effect. It converts kinetic energy of the convection into magnetic energy. The precise nature of these non-dissipative turbulence effects and the  $\alpha$-effect in particular, are still under discussion. 

In solar dynamo models, the toroidal part of the magnetic field is thought to be the dominant one which can be explained by the shearing of any poloidal field by the differential rotation of the Sun ($\Omega$-effect). Since sunspot groups typically appear as bipolar magnetic-field regions, they are often attributed to a toroidal field piercing locally through the solar surface. As for generating a poloidal field from a toroidal one, two of the most promising mechanisms can be described as (i) the effect of rotating stratified turbulence, where helical twisting of the toroidal field lines by the Coriolis force generates a poloidal field (turbulent $\alpha$-effect) \citep{1955ApJ...122..293P,1955ApJ...121..491P,2005PhR...417....1B}, and (ii) the Babcock-Leighton (BL) mechanism \citep{1961ApJ...133..572B,1964ApJ...140.1547L,2014ARA&A..52..251C}.

In the BL mechanism, the surface meridional flow and supergranular diffusion leads to transportation and diffusion of bipolar active regions, which are tilted with regards to the east-west direction. This process is considered as a poloidal magnetic field source at the solar surface \citep{1961ApJ...133..572B,1964ApJ...140.1547L,1989Sci...245..712W,1991ApJ...375..761W}. The poloidal field sources are transported to the solar poles by the poleward meridional flow on the surface and cause the polarity reversal at sunspot maximum \citep{2015Sci...347.1333C}. The meridional flow then penetrates below the base of the convection zone and is responsible for the generation and equatorward propagation of the bipolar activity structures at low latitudes at the solar surface \citep{1999ApJ...518..508D,2001ApJ...551..576N,2002Sci...296.1671N}. The inclusion of a poleward surface meridional flow along with an equatorward deep-seated meridional flow led to the development of the so-called flux transport (FT) dynamo models \citep{1991ApJ...383..431W,1999ApJ...518..508D,2001ApJ...551..576N}. There are several types of FT dynamo models, which produce the poloidal field either via a pure BL-mechanism or a pure $\alpha$-turbulent effect operating in the tachocline, or, alternatively, in the whole convection zone. More recently, dynamo models operating with $\alpha$-turbulence and BL-mechanisms simultaneously as poloidal field sources have also emerged \citep{2001ApJ...559..428D,2014A&A...563A..18P,2013ApJ...779....4B}.

The current mean-field solar dynamo models can reproduce some specific aspects of the long-term modulation of the amplitude of the solar activity and changes in the symmetry of the field topology (parity). There are at least two ways to reproduce long-term modulations in solar mean-field dynamos; via sudden changes in the governing parameters of the solar dynamo \citep{2008SoPh..250..221M,2009RAA.....9..953C,2009SoPh..254..345U,2010ApJ...720.1030C,2010ApJ...724.1021K,2012PhRvL.109q1103C,2013RAA....13.1339K,2013ApJ...777...71O} or via the Lorentz force as a back-reaction of the magnetic fields on the velocity field \citep{1996A&A...307L..21T,1997A&A...322.1007T,1999A&A...343..977K,1999A&A...346..295P,2006ApJ...647..662R,2006MNRAS.371..772B}, which can be two-fold: as a large-scale effect on the differential rotation \citep{1975JFM....67..417M} and through the magnetic suppression of the turbulence, which is a driver of the generation of solar differential rotation ($\Lambda$-effect and $\Lambda$-quenching, e. g. \citet{1994A&A...292..125K}).

In this study, we use a BL-type FT solar dynamo model \citep{2005ApJ...622.1320R,2006ApJ...647..662R} to study the nature and occurrence characteristics of grand minimum and maximum-like periods emerging from simulations involving random fluctuations in either the BL-effect or the generation of the flow system (differential rotation and meridional circulation) and compare them to the results obtained from the cosmogenic isotope records. Section~\ref{sec:dynamo} describes the dynamo model used in this study, while we explain the generation of grand minimum and maximum-like periods in Section~\ref{sec:randomness}. The performed analyses are explained in Section~\ref{sec:analyses} and results from the simulations are presented in Sections~\ref{sec:resultsRL} and~\ref{sec:resultsRB}. Discussion and conclusions are given in Section~\ref{sec:discconc}.

\section{The Dynamo Model} \label{sec:dynamo}

The BL-FT dynamo model in this study uses the mean field differential rotation and meridional circulation model of \citet{2005ApJ...622.1320R} coupled with the axisymmetric mean field induction equation \citep{2006ApJ...647..662R}. The momentum equation for the mean differential rotation and meridional circulation model (the $\Lambda$-mechanism) are updated following \citet{2005AN....326..379K}. While the large-scale flow field alone does not act as a dynamo, it does amplify and advect magnetic fields on the one hand. It is modified, on the other hand, by the action of Lorentz forces exerted by the fields generated. Dynamo action is accomplished by adding a source term mimicking the BL-effect in the induction equation \citep[see][for details]{2006ApJ...647..662R}.
 
Using the angular momentum transport equations given in \citet{2005AN....326..379K} instead of \citet{2005ApJ...622.1320R}'s parameterisations required us to use a different set of parameters than those used in \citet{2005ApJ...622.1320R,2006ApJ...647..662R}. To generate solar-like meridional circulation, differential rotation, and oscillatory magnetic field solutions, the turbulent viscosity ($\nu_{t}$) and the heat conductivity ($\kappa_{t}$), which are assumed to be constant throughout the convective zone, are taken as $2\times 10^8$ m$^2$\,s$^{-1}$. The parameters $\nu_{t}$ and $\kappa_{t}$ modify the meridional flow speed (keeping $\nu_{t}$/$\kappa_{t}$ ratio constant). The value of superadiabaticity in the overshoot region, which affects the differential rotation profile throughout the domain, is taken as $\delta_{\rm os}$=$2\times 10^{-5}$. We also chose the amplitude of the $\alpha$-effect as 0.4 m\,s$^{-1}$. We refer the reader to \citet{2005ApJ...622.1320R,2006ApJ...647..662R} for more detailed information on the dynamo model.

\begin{figure*}[htb!]
\begin{center}
{\includegraphics[width=\textwidth]{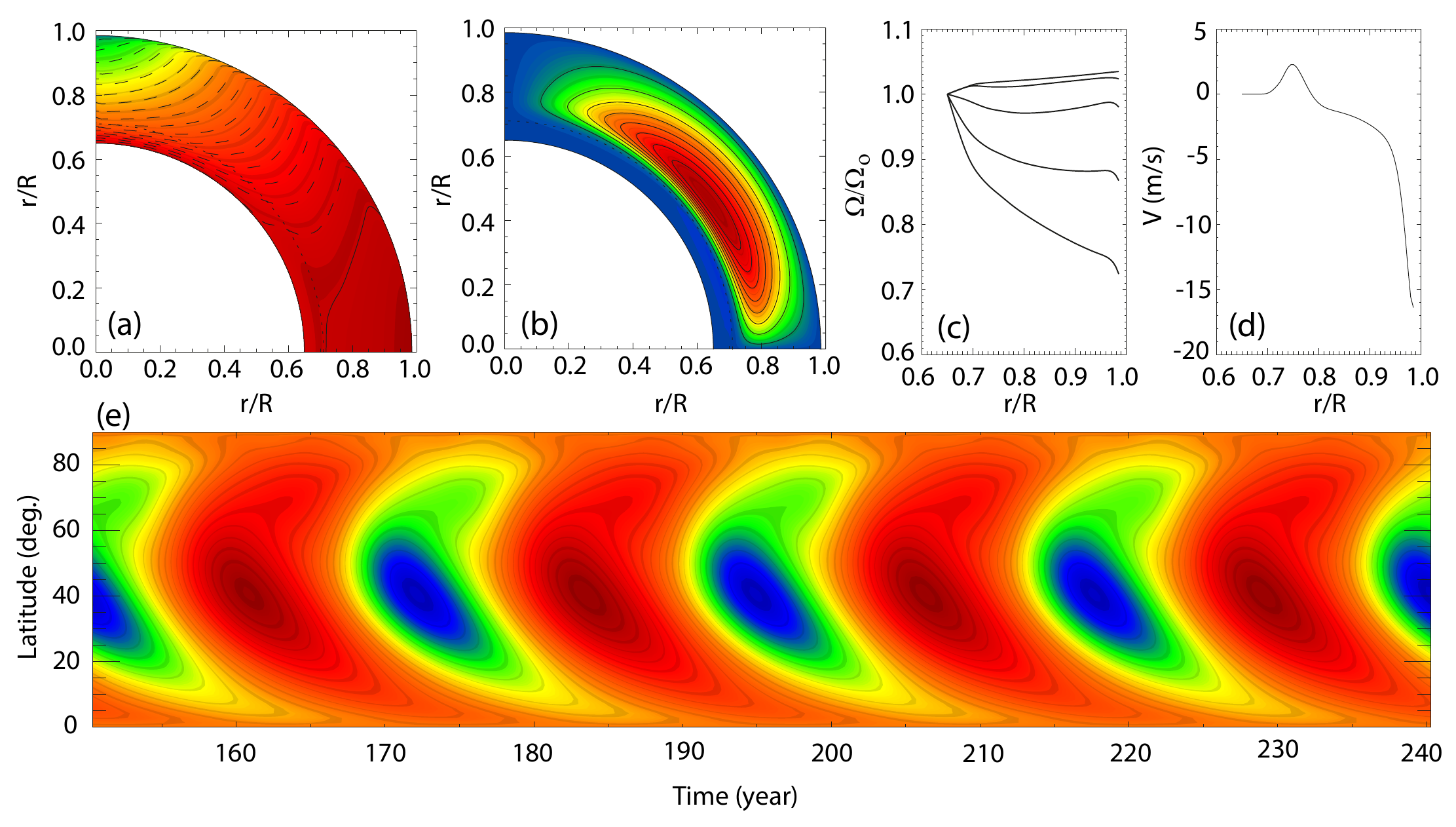}}
\caption{The reference differential rotation (a) and meridional flow (b) contour plots, and their radial profiles, respectively (c, d). The meridional circulation (c) is given at 45$^{\circ}$ latitude, while the radial profile of the differential rotation (c) is given at, from top to bottom, 0$^{\circ}$, 20$^{\circ}$, 40$^{\circ}$, 60$^{\circ}$, and 80$^{\circ}$ latitudes in units of the rotation rate of solar interior, $\Omega_{0}=420$~nHz. The bottom panel (e) shows the butterfly diagram generated at 0.71$R_{\odot}$ for given parameters without any random fluctuations (see text).} 
\label{fig:MCDFB}
\end{center}
\end{figure*}

The computed reference differential rotation and meridional flow radial profiles, their contour plots, and the butterfly diagram calculated at 0.71$R_{\odot}$ are shown in Figure~\ref{fig:MCDFB}. The differential rotation contours and profile, which is given in units of the rotation rate of solar interior, $\Omega_{0}=420$~nHz, show a subsurface shear-layer (Figure~\ref{fig:MCDFB}a and c), which is in agreement with the helioseismic observations \citep{2003ARA&A..41..599T,2009LRSP....6....1H} and the meridional circulation speed on top of the domain at 45$^{\circ}$ latitude is $\sim$16 m\,s$^{-1}$ (Figure~\ref{fig:MCDFB}d). The butterfly diagram obtained from the given parameters show that the solar cycle starts at around 45$^{\circ}$ latitude and the magnetic activity propagates equator-ward and poleward (Figure~\ref{fig:MCDFB}e), resembling the observed sunspot cycles. The average strength of the generated magnetic field in the reference model is around 0.38 Tesla.

\section{Generation of Grand Solar Minimum and Maximum} \label{sec:randomness}

To generate grand solar minimum and maximum-like periods in our simulations, we used random fluctuations in the $\Lambda$-mechanism and in the BL-source term, separately. 

\subsection{Random fluctuations in $\Lambda$-mechanism}

Studies on the surface rotation revealed that it shows variations over a solar cycle, where it is faster during a minimum \citep{2006SoPh..237..365B}. It was shown by helioseismic observations that the 11-year torsional oscillations \citep{1980ApJ...239L..33H}, are not only a surface phenomenon but they extend down to solar convection zone \citep{2000Sci...287.2456H}. These observations provide information on the solar interior and rotation rates from inversions, 1$\sigma$ confidence intervals of which within the convection zone is less than 1\% of the rotation rate \citep{2002ApJ...567.1234S}. 

The average meridional flow at the surface, moving towards to the solar poles, is about 16 to 20 m\,s$^{-1}$. It shows variations of about 10 m\,s$^{-1}$ at the solar surface, and also towards the interior of the Sun with the solar cycle. Amplitude of the meridional circulation at the surface is larger during a solar cycle minimum \citep{2008SoPh..252..235G,2010Sci...327.1350H,2015SoPh..290.3113K}. 

The variations observed in the differential rotation and the meridional flow are mainly caused by the changes in magnetic field strength over a solar cycle. Little is known, however, about short-term variations in these flows, which could be caused by the turbulent Reynolds stress ($\Lambda$-mechanism) and its perturbations \citep{2005ApJ...631.1286R,2007AN....328.1096R}.

In this study, we used the angular momentum transport equations after \citet{2005AN....326..379K}, where the correlation tensor of the fluctuating velocity $\vec{u'}$ gives the angular momentum transport,

\begin{eqnarray}
Q_{ij}=\langle u_{i}^{\prime}(x,t)u_{j}^{\prime}(x,t) \rangle
\label{eq:AngMom1}
\end{eqnarray}

\noindent of which the off-diagonal components $Q_{r,\phi}$ and $Q_{\theta,\phi}$ in spherical coordinates are proportional to,

\begin{eqnarray}
Q_{r\phi}^{\Lambda}=\nu_{\rm T}\Omega V \sin\theta
\label{eq:AngMom2}
\end{eqnarray}

\begin{eqnarray}
Q_{\theta\phi}^{\Lambda}=\nu_{\rm T}\Omega H \cos\theta
\label{eq:AngMom3}
\end{eqnarray}

The parameters $V$ and $H$ denote vertical and horizontal angular momentum fluxes, where $\nu_{\rm T}$ represents the eddy viscosity \citep{2005AN....326..379K}.

The random fluctuations with correlation length- and time-scales added to the radial (equation~\ref{eq:AngMom2}) and latitudinal (equation~\ref{eq:AngMom3}) angular momentum transport equations as multipliers,

\begin{eqnarray}
{\rm RN}_{r\phi}^{\Lambda}=\left[1+\frac{c\zeta_{r}(r,\theta)}{\sigma_{r}}\right]
\label{eq:AngMom2RND1}
\end{eqnarray}

\begin{eqnarray}
{\rm RN}_{\theta\phi}^{\Lambda}=\left[1+\frac{c\zeta_{\phi}(r,\theta)}{\sigma_{\theta}}\right]
\label{eq:AngMom2RND2}
\end{eqnarray}

The parameters $\zeta_{r}(r,\theta)$ and $\zeta_{\phi}(r,\theta)$ represent random functions in radial and latitudinal directions, where $c$ denotes the amplitude of the random fluctuations with respect to the mean \citep[see][for details]{2005ApJ...631.1286R}. Free parameters of the applied random fluctuations are: (i) the amplitude of the random fluctuation $c$, (ii) the correlation length-scales in radius and latitude ($\Delta r$ and $\Delta\theta$ incorporated in the $\zeta$-functions), and (iii) the correlation time-scale $\tau_{\rm c}$. The quantity $\Delta$ measures the length-scale in radial and latitudinal direction in  units of the domain size, while $\tau_{\rm c}$ measures the time-scale in units of the inverse of the rotation rate of the solar interior ($\Omega_{0}^{-1}$). The correlation time-scale $\tau_{\rm c}$ is introduced in the random function $\zeta$ \citep[equation 3 in][]{2005ApJ...631.1286R}. We must note that our random fluctuations are added as multipliers to the angular momentum transport equations, meaning that they are real-signal dependent and this might not always be the case.

In this study, to generate random fluctuations in the $\Lambda$-mechanism, we choose different amplitudes of $c$ such as, 0.1, 0.2, and 0.3, which provide us with fluctuations below 1$\sigma$ level. Additionally, the correlation time-scales are 4\,$\Omega_{0}^{-1}$ and 16\,$\Omega_{0}^{-1}$ corresponding roughly to 16 and 64 days, respectively. Furthermore, the correlation length-scales are taken as 5\%, 10\%, and 20\% of the thickness of the convection zone (Table~\ref{tab:RNLambda}).

\begin{table}[htb!]
\centering
\caption{The parameters used to generate random fluctuations in the $\Lambda$-mechanism. The parameter $\tau_{\rm c}=4$\,$\Omega_{0}^{-1}$$\approx16$ days, while $\tau_{\rm c}=16$\,$\Omega_{0}^{-1}$$\approx64$ days.}
\begin{tabular}{lcccccc}
\hline \hline
E 	&$c$		&$\tau_{\rm c}$ ($\Omega_{0}^{-1}$)	&$\Delta$		&Dur.(yr)	&$\#$GMin	&$\#$GMax			\\ \hline
1			&	0.1	&	4.0		&	0.05				& 7936		& --		& --				\\	
2			&	0.2	&	4.0		&	0.05				& 7857		& --		& --				\\	
{\bf 3}		&{\bf 0.3}	&{\bf	4.0}		&{\bf	0.05}				&{\bf 7857}	&{\bf 45}	&{\bf 29}				\\
4			&	0.1	&	16.0		&	0.05				& 7857		& --		& --				\\
{\bf 5}		&{\bf	0.2}	&{\bf	16.0}		&{\bf	0.05}				&{\bf 7857}	&{\bf 38}	&{\bf 35}				\\
{\bf 6}		&{\bf	0.3}	&{\bf	16.0}		&{\bf	0.05}				&{\bf 7857}	&{\bf 30}	&{\bf 35}				\\
7			&	0.1	&	4.0		&	0.1				& 7857		& --		& --				\\
8			&	0.2	&	4.0		&	0.1				& 7857		& --		& --				\\
{\bf 9}		&{\bf	0.3}	&{\bf	4.0}		&{\bf	0.1}				&{\bf 7858}	&{\bf 33}	&{\bf 38}				\\
10			&	0.1	&	16.0		&	0.1				& 7858		& --		& --				\\
{\bf 11}		&{\bf	0.2}	&{\bf	16.0}		&{\bf	0.1}				&{\bf 7858}	&{\bf 35}	&{\bf 34}				\\
{\bf 12}		&{\bf	0.3}	&{\bf	16.0}		&{\bf	0.1}				&{\bf 7857}	&{\bf 31}	&{\bf 34}				\\
13			&	0.1	&	4.0		&	0.2				& 7884		& --		& --				\\
14			&	0.2	&	4.0		&	0.2				& 7903		& --		& --				\\
{\bf 15}		&{\bf	0.3}	&{\bf	4.0}		&{\bf	0.2}				&{\bf 7887}	&{\bf 36}	&{\bf 34}				\\
16			&	0.1	&	16.0		&	0.2				& 7862		& --		& --				\\
{\bf 17}		&{\bf	0.2}	&{\bf	16.0}		&{\bf	0.2}				&{\bf 7940}	&{\bf 34}	&{\bf 30}				\\
{\bf 18}		&{\bf	0.3}	&{\bf	16.0}		&{\bf	0.2}				&{\bf 7874}	&{\bf 34}	&{\bf 34}				\\ \hline 
\end{tabular}
\label{tab:RNLambda}
\end{table}

\subsection{Random fluctuations in the Babcock-Leighton mechanism}

The Babcock-Leighton mechanism relies on the tilt of bipolar active regions, providing a net transport of magnetic flux of the following polarity to the poles. This net transport is the outcome of competing diffusion and advection of the two polarities \citep{1961ApJ...133..572B,1964ApJ...140.1547L}. The sunspot observations reveal that the average tilt of the bipolar magnetic regions with regards to the east-west direction is around +5$^{\circ}$, however the distribution of these tilt angles are very broad and extends from $-90^{\circ}$ to $+90^{\circ}$ with a full width at half-maximum (FWHM) of typically $30^\circ$--$40^\circ$ \citep{1996ARA&A..34...75H,2016AdSpR..58.1468S}. Random fluctuations are therefore inherent in the Babcock-Leighton mechanism.

We added random fluctuations to the BL-source term \citep[the last term in the right hand side of equation 7 of][]{2006ApJ...647..662R} by multiplying it with  

\begin{eqnarray}
{\rm RN}^{\rm S}=\left[1+\frac{c\zeta_{\rm S}}{\sigma_{\rm S}}\right]
\label{eq:SourceRND}
\end{eqnarray}

\noindent where $c$ is the amplitude of the random fluctuations with respect to the mean and $\zeta_{\rm S}$ represents the random function in the source term. The random fluctuations in the BL-mechanism are also real-signal dependent. The free parameters that are used to generate random fluctuations in the BL-mechanism are given in Table~\ref{tab:RNBL}. Similar to those in Table~\ref{tab:RNLambda}, the correlation time-scales are again chosen as 4\,$\Omega_{0}^{-1}$ and 16\,$\Omega_{0}^{-1}$ corresponding roughly to 16 and 64 days, respectively, and the amplitudes change from 0.1 to 1.0, corresponding to fluctuations below around 3$\sigma$.

\begin{table}[htb!]
\centering
\caption{The parameters used to generate random fluctuations in the BL-mechanism. The parameter $\tau_{\rm c}=4$\,$\Omega_{0}^{-1}$$\approx16$ days, while $\tau_{\rm c}=16$\,$\Omega_{0}^{-1}$$\approx64$ days.}
\begin{tabular}{lccccc}
\hline \hline
E	&c		&$\tau_{\rm c}$ ($\Omega_{0}^{-1}$)	&Dur.(yr)	&$\#$GMin	&$\#$Gmax		\\ \hline
19	&	0.1	&	4.0						& 7907		& --		& --			\\ 
20	&	0.2	&	4.0						& 7882		& --		& --			\\
21	&	0.3	&	4.0						& 7862		& --		& --			\\
22	&	0.4	&	4.0						& 7960		& --		& --			\\
23	&	0.5	&	4.0						& 7998		& --		& --			\\
24	&	0.6	&	4.0						& 7886		& --		& --			\\
25	&	0.7	&	4.0						& 8021		& --		& --			\\
26	&	0.8	&	4.0						& 7928		& --		& --			\\
27	&	0.9	&	4.0						& 7881		& --		& --			\\
{\bf 28}	&{\bf	1.0}	&{\bf	4.0}					&{\bf 7906}	&{\bf 34}	&{\bf 28}			\\
29	&	0.1	&	16.0						& 7895		& --		& --			\\
30	&	0.2	&	16.0						& 7892		& --		& --			\\
31	&	0.3	&	16.0						& 7981		& --		& --			\\
32	&	0.4	&	16.0						& 8216		& --		& --			\\
{\bf 33}	&{\bf	0.5}	&{\bf	16.0}					&{\bf 8122	}	&{\bf 35}	&{\bf 22}			\\
{\bf 34}	&{\bf	0.6}	&{\bf	16.0}					&{\bf 7909	}	&{\bf 28}	&{\bf 24}			\\
{\bf 35}	&{\bf	0.7}	&{\bf	16.0}					&{\bf 7901	}	&{\bf 34}	&{\bf 31}			\\
{\bf 36}	&{\bf	0.8}	&{\bf	16.0}					&{\bf 7969	}	&{\bf 35}	&{\bf 31}			\\
{\bf 37}	&{\bf	0.9}	&{\bf	16.0}					&{\bf 7885	}	&{\bf 40}	&{\bf 33}			\\
{\bf 38}	&{\bf	1.0}	&{\bf	16.0}					&{\bf 7914	}	&{\bf 37}	&{\bf 35}			\\ \hline
\end{tabular}
\label{tab:RNBL}
\end{table}

\section{Analyses} \label{sec:analyses}

The basic periodicity in each simulation is around 11-year, closely resembling the sunspot cycle. Before performing any further analyses, we first smoothed each $B_\phi$ calculated as the absolute maximum value at all latitudes at 0.71$R_{\odot}$ using a Butterworth filter of degree 5 with a cutoff frequency of (23 yr)$^{-1}$ to obtain similar features in frequency domain as the cosmogenic nuclides do, which are used in \citet{2015A&A...577A..20I,2016SoPh..291..303I}. Following to that, we truncated the first $\sim 150$ years of each simulation to omit the growth period of the dynamo in the analyses.

\subsection{Identification of grand minima and maxima}\label{sec:analysesIdent}

To identify the grand minimum and maximum periods in the simulations, we first checked the range of change (${\rm max}B_\phi-{\rm min}B_\phi)/{\rm min}B_\phi$ in the calculated magnetic field strengths for each simulation. The simulations that show range of change values $\geq1$ are selected for further analyses, whereas those showing range of change values $<1$ are disregarded. We then subtract the mean values from each selected simulation to use the zero-crossing method to determine the start and the end dates of the peaks and dips throughout the time. To identify grand minimum and maximum periods, we used $\pm$1.645$\sigma$ variation around the mean as a threshold value, meaning that a data point arbitrarily chosen in the time-series will fall within $\pm$1.645$\sigma$ around the mean with a 90\% probability. A period with a smaller (greater) amplitude than the defined threshold value will be considered as a grand minimum (maximum) candidate. If the duration of this candidate is longer than 22 yr (more than twice of the basic 11-year cycle), then the event is identified as a grand minimum or a grand maximum period.

\subsection{Waiting Time Distribution}

For the waiting time distribution analyses, which is also known as inter-arrival time distribution, we first define the waiting times as the time span between the minimum (maximum) times of two consecutive minima (maxima) events. The complementary cumulative probability distribution of the waiting times between discrete events has been broadly used in physical sciences to investigate whether the occurrence of these events reflect random or time-dependent, memory-bearing processes \citep{2000ApJ...536L.109W,2001ApJ...555L.133L,2003SoPh..214..361W}. A purely random Poisson process, which does not include a memory effect, is represented by an exponential waiting time distribution (WTD), where the occurrence of an event is independent of the preceding event \citep{2007A&A...471..301U}. On the contrary, a power-law WTD points to a memory-bearing process, where the occurrence is dependent on the previous event \citep{Clauset2009}. 

The complementary cumulative distribution function is defined as the probability $L$ that an event $X$ with a certain probability distribution $l(x)$ will be found at a value more than or equal to $x$ \citep{Clauset2009,Guerriero2012},

\begin{equation}
{L~(X \ge x)}=1-\int_{-\infty}^{~x}{l(x)~dx}=\int_{~x}^{~\infty}{l(x)~dx}
\label{eq:CCDF}
\end{equation}

Following this step, we fit a power-law (Eq.~\ref{eq:PL}) \citep{Clauset2009} and an exponential distribution (Eq.~\ref{eq:EL}) using the maximum likelihood method (MLM). The MLM is robust and accurate for estimation of the parameters of the distributions we consider here \citep{Clauset2009,Guerriero2012}. 

\begin{equation}
{L(x)} \propto {x^{-\alpha}}
\label{eq:PL}
\end{equation}

\begin{equation}
{L(x)} \propto {{\rm exp}\left(\frac{-x} {\tau}\right)} 
\label{eq:EL}
\end{equation}

\noindent where $\alpha$ and $\tau$ indicate the scaling and the survival parameters of the power-law and exponential probability distributions, respectively \citep{VirkarandClauset2014}.

The goodness of the fits are calculated using two-sample Kolmogorov-Smirnov (KS) tests. To achieve this, we generated 1000 data sets for exponential and power law distributions, separately, using the calculated scaling ($\alpha$) and survival ($\tau$) parameters. We then compare each generated data to the underlying data with the two-sample KS test. Using the distributions of the p-values obtained from the two-sample KS test, we decide whether the distribution of the underlying WTD is better represented with a exponential or a power-law distribution. Larger p-values implies better representation of the data.

\subsection{Variations in meridional circulation and radial differential rotation}

We used a two-sample KS test to investigate whether the meridional circulation and the differential rotation during the identified grand minima in our simulations are statistically different from those observed during the grand maxima periods. To achieve these objectives, we first filtered the data sets using a Butterworth filter of degree 5 with a cutoff frequency of (23 yr)$^{-1}$, as it was used for the simulated magnetic field strength data. Following to this, we isolated the meridional circulation rates, which are calculated at 0.985$R_{\odot}$ and at 45$^{\circ}$ latitude (hereafter MC), during the grand minima and maxima periods, separately. For the differential rotation, we used two approaches that are (i) the difference between the rotation values calculated at 0.70$R_{\odot}$ and 0.985$R_{\odot}$, and at 45$^{\circ}$ latitude (hereafter $\Delta{\rm DR_{Rad}}$), and (ii) the difference between the rotation values calculated at 0$^{\circ}$ and 60$^{\circ}$ latitudes and at 0.985$R_{\odot}$ (hereafter $\Delta{\rm DR_{Lat}}$). Similar to the procedure used for the meridional circulation, we also isolated the differential rotation values observed during the grand minima and maxima periods, separately, in our simulations. 

\begin{figure*}[htb!]
\begin{center}
{\includegraphics[width=\textwidth]{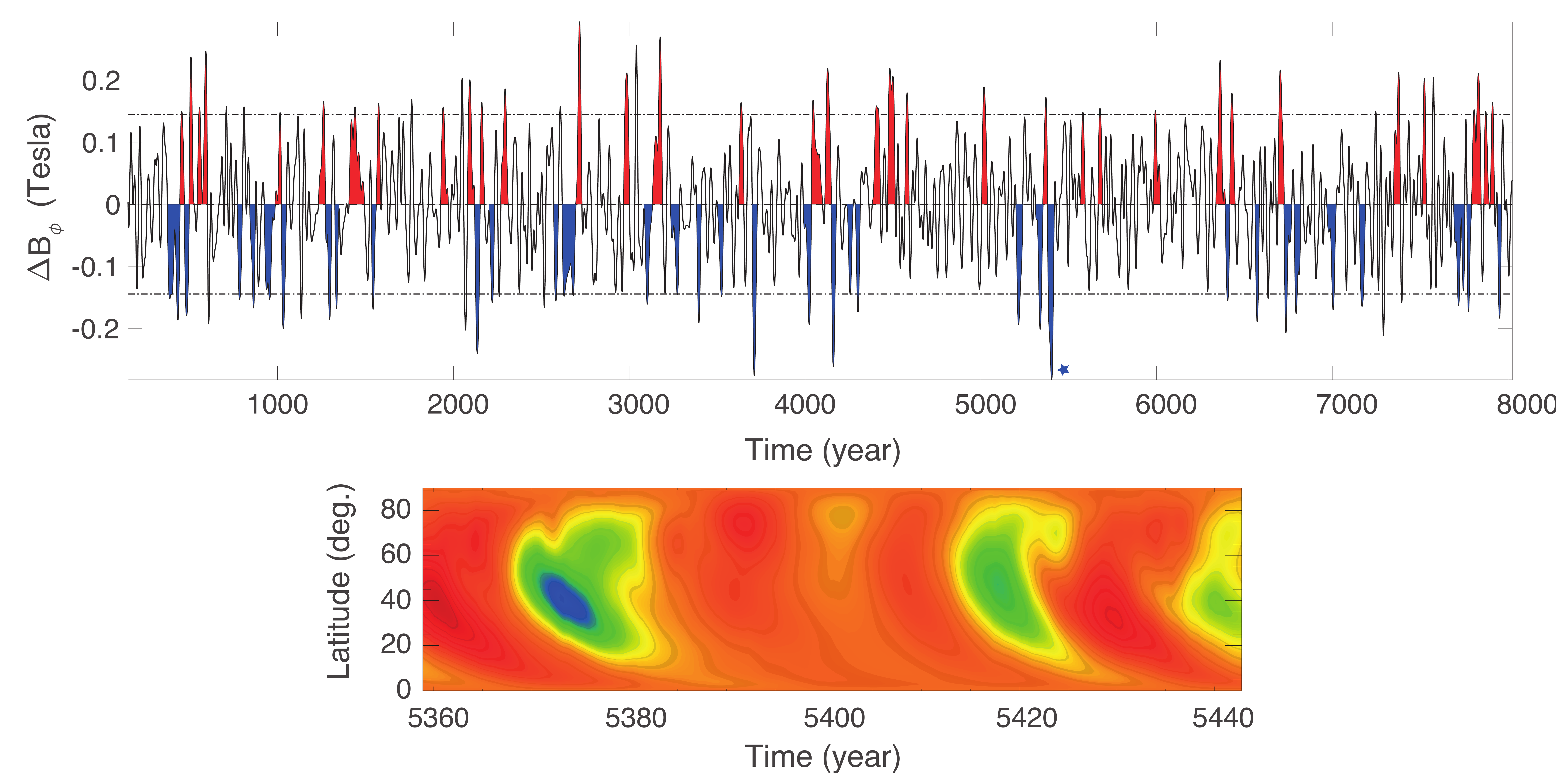}}
\caption{The top panel shows the maximum magnetic field strength ($\Delta B_{\phi}$) at 0.71$R_{\odot}$ of experiment 18 (Table~\ref{tab:RNLambda}). The blue and red fill colours represent grand minima and maxima, respectively, while the dashed lines show the calculated threshold values ($\pm$1.645$\sigma$, see section~\ref{sec:analysesIdent}). The bottom panel shows the butterfly diagram of $B_\phi$ at 0.71$R_{\odot}$ for the grand minimum period between $\sim5380$--5420, marked with a blue star in the top panel.} 
\label{fig:E18MinMax}
\end{center}
\end{figure*}

\section{Results from random fluctuations in $\Lambda$-mechanism} \label{sec:resultsRL}

\subsection{Identification of grand minima and maxima}

We identified grand minimum and maximum events in our simulations, in which random fluctuations in the $\Lambda$-mechanism are used, via the criteria defined in this study. Simulations, which show grand minima and maxima are shown in bold face in Table~\ref{tab:RNLambda}. For the correlation time-scale of $\sim16$ days ($\tau_{\rm c}=4$\,$\Omega_{0}^{-1}$), only the fluctuations that have the amplitude of 0.3 at all correlation length-scales (0.05, 0.1, and 0.2) show grand minimum and maximum periods, while for the correlation time-scales of $\sim64$ days ($\tau_{\rm c}=16$\,$\Omega_{0}^{-1}$) fluctuations, which have the amplitudes of 0.2 and 0.3 show grand minimum and maximum periods at all correlation length-scales (Table~\ref{tab:RNLambda}). The results imply that the amplitudes of random fluctuations smaller than 33\% of the mean, which corresponds to the Reynolds stress with $\leq1\sigma$ fluctuations, are able to generate grand minima and maxima at correlation length-scales of 0.05, 0.1, and 0.2 in our simulations.

\begin{table*}[htb!]
\centering
\caption{Results for the calculated power-law and exponential fits for the WTDs of grand minima and maxima identified in each simulation, where random fluctuations in the $\Lambda$-mechanism is used (see text).}
\begin{tabular}{lcccccccccc}
\hline \hline
\multicolumn{1}{c}{} & \multicolumn{5}{c}{Grand Min} & \multicolumn{5}{c}{Grand Max} \\

E 	 &$p_{\rm pow}$ 	&$p_{\rm exp}$ 	&Time (\%)	&Dist.	&Dur. (yr)  &$p_{\rm pow}$ 	&$p_{\rm exp}$ 	& Time (\%)	&Dist.	&Dur. (yr)		\\ 
\hline
3	&0.04		&0.08		&18			&Exp.	&32		&0.07		&0.41		&15			&Exp.	&40	\\
5	&0.09		&0.19		&16			&Exp.	&32		&0.01		&0.48		&15			&Exp.	&33	\\
6	&0.00		&0.06		&12		 	&Exp.	&32		&0.20		&0.43		&14			&Exp.	&31	\\
{\bf 9}	&{\bf 0.25}		&{\bf 0.23}		&{\bf 13}			&{\bf Pow.}	&{\bf 31}		&{\bf 0.51}		&{\bf 0.15}		&{\bf 17}			&{\bf Pow.}	&{\bf 35}	\\
{\bf 11}	&0.02		&0.14		&15				&Exp.		&35			&{\bf 0.41}		&{\bf 0.28}		&{\bf 15}			&{\bf Pow.}	&{\bf 35}	\\
12	&0.40		&0.43		&13			&Exp.	&32		&0.04		&0.45		&17			&Exp.	&38	\\
15	&0.17		&0.18		&14			&Exp.	&31		&0.14		&0.22		&16			&Exp.	&37	\\
17	&0.05		&0.08		&14			&Exp.	&34		&0.22		&0.32		&15			&Exp.	&39	\\
18	&0.21		&0.34		&16			&Exp.	&37		&0.08		&0.42		&16			&Exp.	&37	\\
\hline
\end{tabular}
\label{tab:gminmaxLambda}
\end{table*}

An example to the identified grand minimum and maximum periods in the temporal evolution of experiment 18 (Table~\ref{tab:RNLambda}), which are generated by applying random fluctuations in the $\Lambda$-mechanism, is represented in Figure~\ref{fig:E18MinMax}. The identified grand minima and maxima are shown as blue and red filled periods, respectively, while the dashed-lines show the calculated threshold values based on the $\pm$1.645$\sigma$ deviation around the mean. We also show a butterfly diagram of $B_\phi$ at 0.71$R_{\odot}$ for the identified minimum period between $\sim5380$--5420 (the bottom panel of Figure~\ref{fig:E18MinMax}). The strength of the toroidal field during this period is much smaller than before entering the minimum and slowly increases towards the termination of the minimum.

\subsection{Waiting time distributions and durations of grand minima and maxima}

Following the identification of grand minima and maxima in our simulations, we performed WTD analyses to investigate whether they show an exponential or a power-law distribution, meaning that whether the $\leq1\sigma$ fluctuations in the $\Lambda$-mechanism are translated into the magnetic field as a purely random or a memory-bearing signal, respectively. Resulting $p_{\rm pow}$, and $p_{\rm exp}$ values that indicate the WTDs of grand minima and maxima are better represented either by a power-law or an exponential distribution are given in Table~\ref{tab:gminmaxLambda}.

\begin{figure*}[htb!]
\begin{center}
{\includegraphics[width=5.2in]{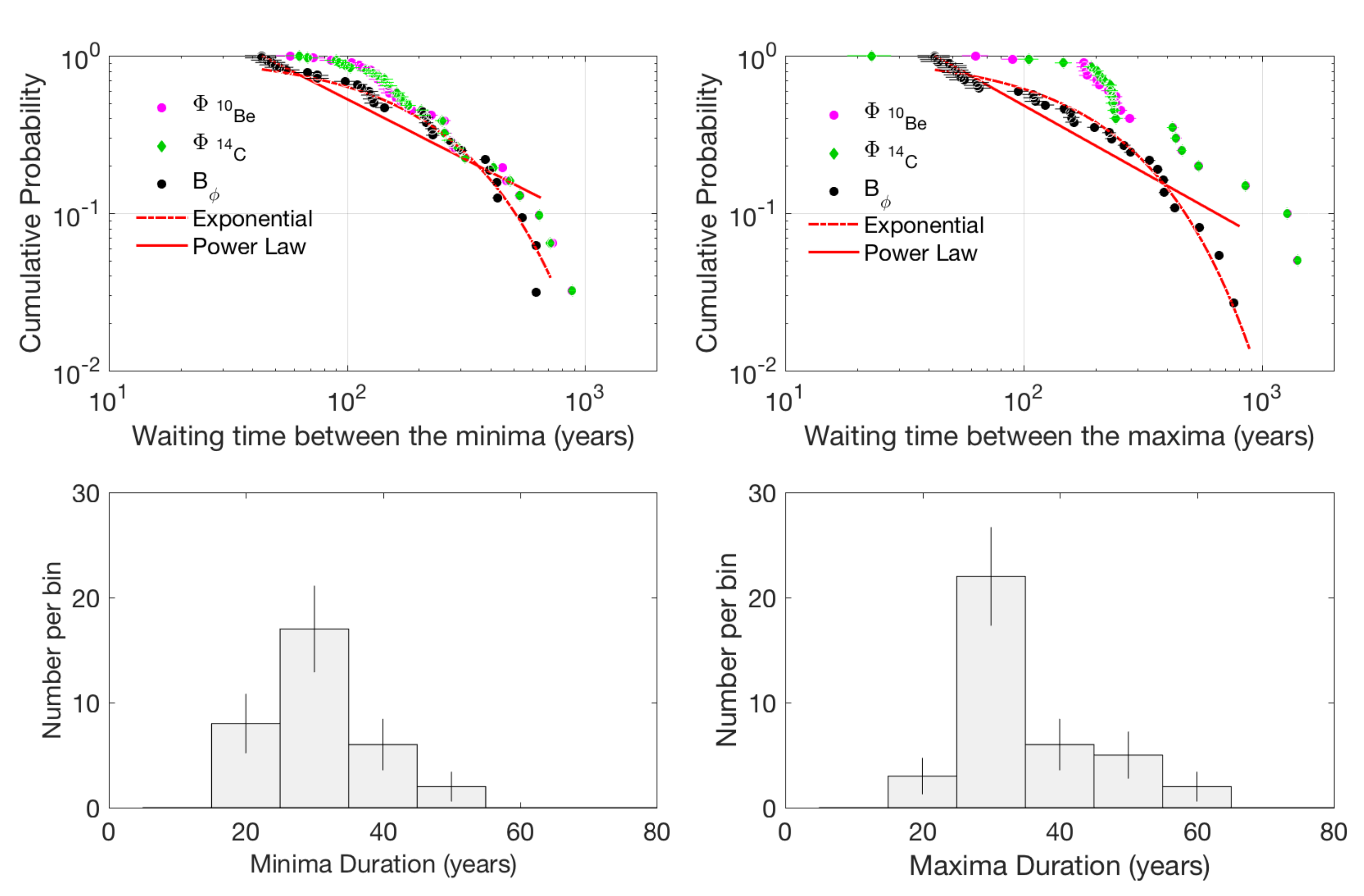}}
\caption{The top panel shows the WTD of grand minima and maxima from experiment 9 together with the observed data points from $^{10}$Be and $^{14}$C based solar modulation potentials from \citet{2015A&A...577A..20I}. The bottom panel shows the distribution of the durations of grand minima and maxima.} 
\label{fig:E9WTDDur}
\end{center}
\end{figure*}

The results show that the majority of the WTDs are best represented by an exponential distribution, except from the WTDs of grand minima and maxima of experiment 9, and the WTDs of grand maxima of experiment 11 (Table~\ref{tab:gminmaxLambda}). It must be noted that the $p_{\rm pow}$, and $p_{\rm exp}$ values for the WTDs of grand minima from experiment 9 differs at the 1\% level. The WTDs of experiment 9 are shown in Figure~\ref{fig:E9WTDDur} together with the WTDs of grand solar minima and maxima from simultaneous changes in $^{10}$Be and $^{14}$C based solar modulation potentials, which were proposed to show power-law distributions \citep{2015A&A...577A..20I}. Here, we also show the distributions of the durations of grand minimum and maximum periods found in experiment 9.

The results for all the experiments show that the distributions of durations of grand minima and maxima are of log-normal distributions, and the average durations are clustered around $\sim33$ years for grand minima and $\sim36$ year for grand maxima (Table~\ref{tab:gminmaxLambda}).

\begin{figure*}[htb!]
\begin{center}
{\includegraphics[width=5.5in]{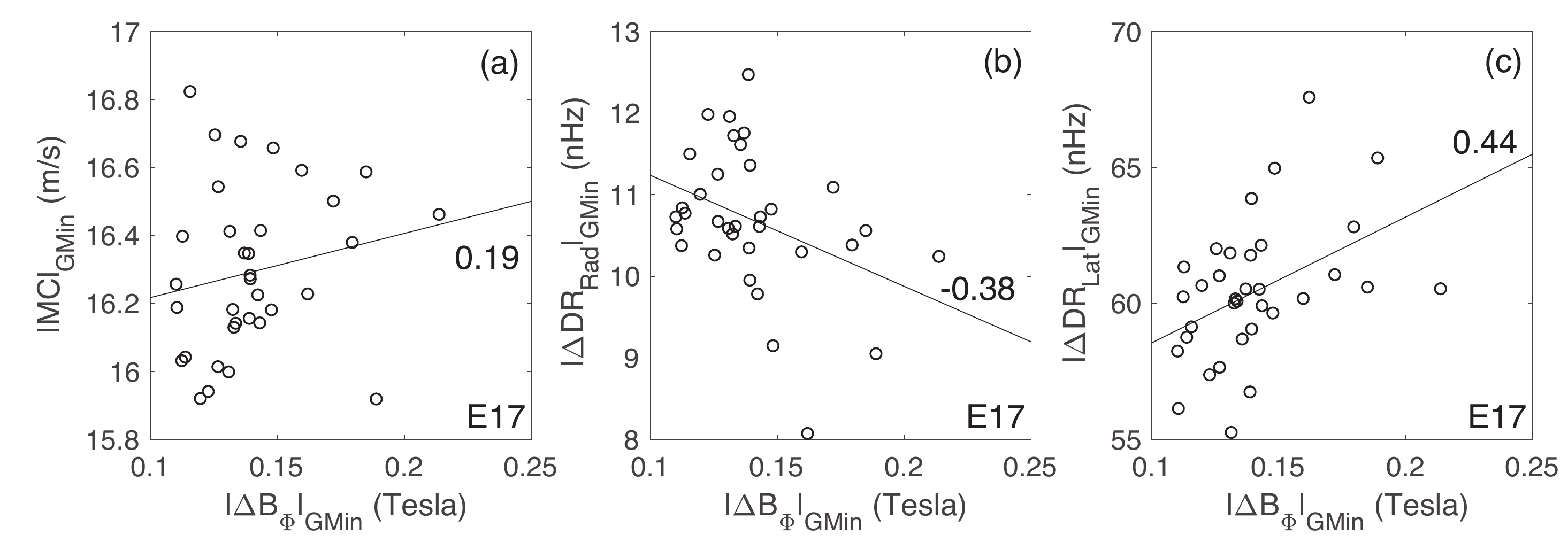}}
\caption{The panel (a) shows the correlation between the amplitudes of surface meridional circulation at 45$^{\circ}$ latitude and the amplitudes of grand minima for experiment 17 (R=0.19). The panel (b) shows the correlation between the amplitudes of difference in rotation rates between 0.70$R_{\odot}$ and 0.985$R_{\odot}$, and at 45$^{\circ}$ latitude ($\Delta{\rm DR_{Rad}}$) and the amplitudes of grand minima (R=-0.38), while the panel (c) shows the correlation between the amplitudes of difference in rotation rates between 0$^{\circ}$ and 60$^{\circ}$ latitudes and at 0.985$R_{\odot}$ ($\Delta{\rm DR_{Lat}}$) and the amplitudes of grand minima for experiment 17 (R=0.44).}
\label{fig:DRMCRL}
\end{center}
\end{figure*}

\subsection{Meridional circulation and radial differential rotation}

The results from the two sample KS tests showed that the meridional circulation during the grand minimum periods are statistically different than those during the grand maximum periods at 99\% significance level. The results also showed that the meridional circulation during the grand minimum periods tends to be faster than those during the grand maxima. The value of $k_{\rm MC}$, shown in Table~\ref{tab:2KSRLam}, is the test statistic for the two-sample KS test and it indicates the degree of the difference between the MC values during grand minimum and maximum periods. The $k_{\rm MC}$ values of the experiments show that the greatest difference in the MC during grand minima and maxima is observed for experiment 17, while experiment 9 has the smallest difference in MC values.

\begin{table}[htb!]
\centering
\caption{The test statistics values for the two sample KS test for meridional circulation at 45$^{\circ}$ latitude ($k_{\rm MC}$), and the difference in rotation rates between 0.70$R_{\odot}$ and 0.985$R_{\odot}$, and at 45$^{\circ}$ latitude ($k_{\Delta {\rm DR_{Rad}}}$) and the difference in rotation rates between 0$^{\circ}$ and 60$^{\circ}$ latitudes and at 0.985$R_{\odot}$ ($k_{\Delta{\rm DR_{Lat}}}$). Larger $k$ values represent larger differences.}
\begin{tabular}{lccc}
\hline \hline
E	&	$k_{\rm MC}$	&	$k_{\Delta{\rm DR_{Rad}}}$	& $k_{\Delta{\rm DR_{Lat}}}$	\\ \hline
3	&	0.17			&	0.56						& 0.34					\\
5	&	0.20			&	0.48						& 0.23					\\
6	&	0.35			&	0.47						& 0.28					\\
9	&	0.16			&	0.53						& 0.40					\\
11	&	0.38			&	0.47						& 0.31					\\
12	&	0.33			&	0.43						& 0.35					\\
15	&	0.23			&	0.31						& 0.23					\\
17	&	0.45			&	0.36						& 0.22					\\
18	&	0.39			&	0.32						& 0.19					\\
\hline
\end{tabular}
\label{tab:2KSRLam}
\end{table}

\begin{figure*}[htb!]
\begin{center}
{\includegraphics[width=\textwidth]{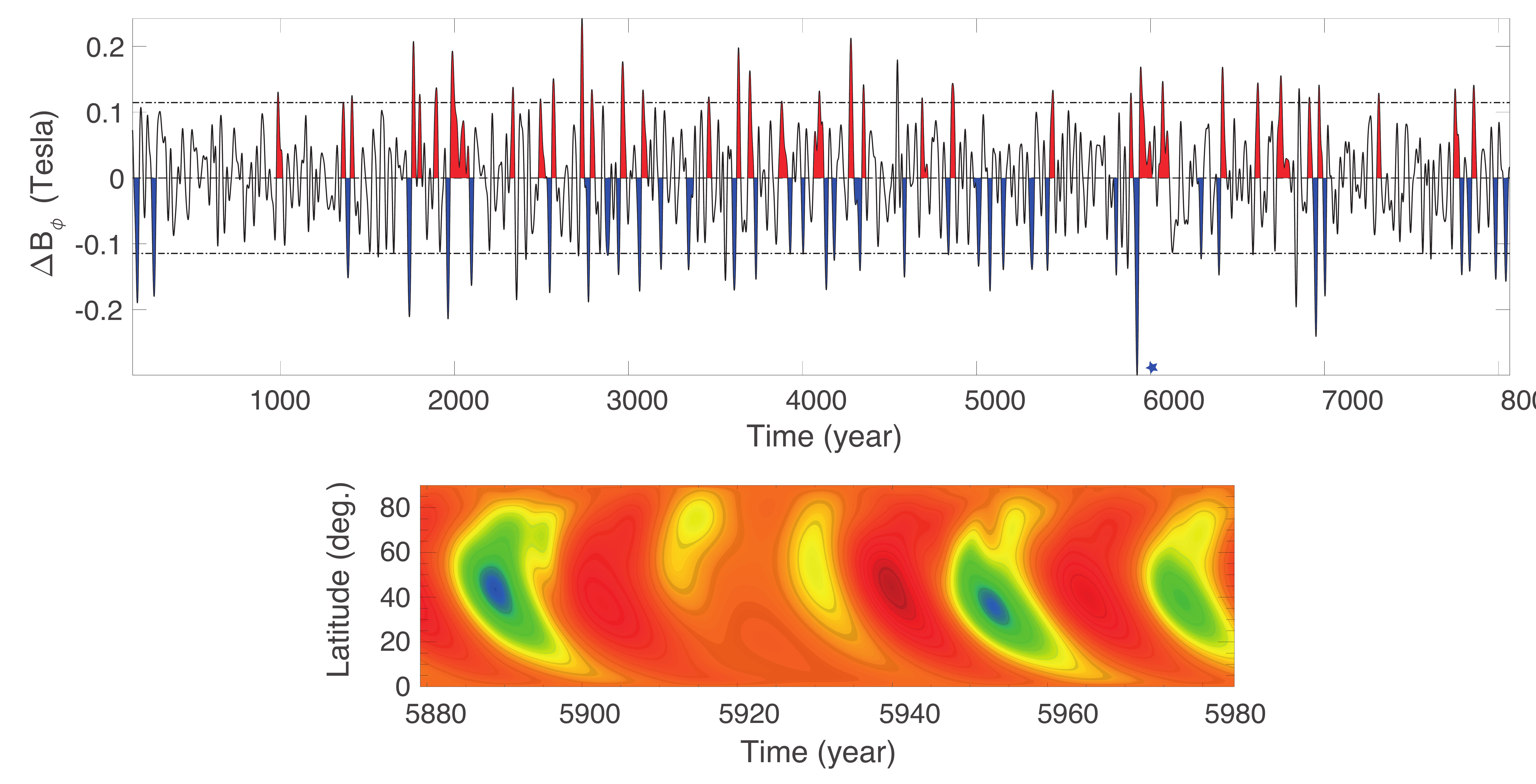}}
\caption{The top panel shows the maximum magnetic field strength ($\Delta B_{\phi}$) at 0.71$R_{\odot}$ of experiment 38 (Table~\ref{tab:RNBL}). The blue and red fill colours represent grand minima and maxima, respectively, while the dashed lines show the calculated threshold values ($\pm$1.645$\sigma$, see section~\ref{sec:analysesIdent}). The bottom panel shows the butterfly diagram of $B_\phi$ at 0.71$R_{\odot}$ for the grand minimum period between $\sim5890$--5940, marked with a blue star in the top panel.} 
\label{fig:E38MinMax}
\end{center}
\end{figure*}

\begin{table*}[htb!]
\centering
\caption{Results for the calculated power-law and exponential fits for the WTDs of grand minima and maxima identified in each simulation, where random fluctuations in the BL-mechanism is used (see text).}
\begin{tabular}{lcccccccccc}
\hline \hline
\multicolumn{1}{c}{} & \multicolumn{5}{c}{Grand Min} & \multicolumn{5}{c}{Grand Max} \\

E 	 &$p_{\rm pow}$ 	&$p_{\rm exp}$ 	&Dist.	&Time (\%)	&Dur. (yr)  &$p_{\rm pow}$ 	&$p_{\rm exp}$ 	&Dist.	& Time (\%)	&Dur. (yr)		\\ 
\hline
28	&0.11		&0.24		&Exp.	&12			&27		&0.04		&0.50		&Exp.	&11			&30	\\
33	&0.13		&0.16		&Exp.	&12			&29		&0.03		&0.30		&Exp.	&9			&34	\\
34	&0.01		&0.11		&Exp.	&10			&28		&0.14		&0.83		&Exp.	&11			&36	\\
{\bf 35}	&{\bf 0.33}		&{\bf 0.25}		&{\bf Pow.}	&{\bf 13}			&{\bf 29}		&0.22		&0.52		&Exp.	&13			&34	\\
36	&0.15		&0.39		&Exp.	&13			&29		&0.14		&0.55		&Exp.	&13			&32	\\
{\bf 37}	&{\bf 0.30}		&{\bf 0.15}		&{\bf Pow.}	&{\bf 14}			&{\bf 28}		&0.16		&0.45		&Exp.	&15			&35	\\
{\bf 38}	&{\bf 0.13}		&{\bf 0.13}		&{\bf Inc.}		&{\bf 14}			&{\bf 29}		&0.08		&0.27		&Exp.	&17			&37	\\
\hline
\end{tabular}
\label{tab:gminmaxBL}
\end{table*}

The two sample KS tests revealed that the $\Delta{\rm DR_{Rad}}$ and the $\Delta{\rm DR_{Lat}}$ values during the grand minima and maxima are statistically different at 99\% significance level. 

The difference in rotation between 0.70$R_{\odot}$ and 0.985$R_{\odot}$, and at 45$^{\circ}$ latitude is larger during grand maxima, while it is smaller during the grand minima. The $k_{\Delta {\rm DR_{Rad}}}$ and $k_{\Delta {\rm DR_{Lat}}}$ values indicates the degree of the difference between the $\Delta{\rm DR_{Rad}}$ and the $\Delta{\rm DR_{Lat}}$ values during grand minimum and maximum periods (Table~\ref{tab:2KSRLam}). The values of $k_{\Delta{\rm DR_{Rad}}}$ calculated for the experiments show that experiment 15 shows the smallest $\Delta{\rm DR_{Rad}}$ between grand minimum and maximum periods, while experiment 3 has the greatest difference. 

The difference between the rotation values calculated at 0$^{\circ}$ and 60$^{\circ}$ latitudes and at 0.985$R_{\odot}$ is larger during grand minima, while it is smaller during grand maxima. Experiment 9 shows the greatest difference in $\Delta{\rm DR_{Lat}}$, while experiment 18 shows the smallest (Table~\ref{tab:2KSRLam}).

Figure~\ref{fig:DRMCRL}a, b, and c show the relationship between the amplitudes of surface meridional circulation at 45$^{\circ}$ latitude and the amplitudes of grand minima (R=0.19), between the amplitudes of difference in rotation rates between 0.70$R_{\odot}$ and 0.985$R_{\odot}$, and at 45$^{\circ}$ latitude ($\Delta{\rm DR_{Rad}}$) and the amplitudes of grand minima (R=-0.38), and between the amplitudes of difference in rotation rates between 0$^{\circ}$ and 60$^{\circ}$ latitudes and at 0.985$R_{\odot}$ ($\Delta{\rm DR_{Lat}}$) and the amplitudes of grand minima for experiment 17 (R=0.44).

\section{Results from random fluctuations in Babcock-Leighton mechanism}
\label{sec:resultsRB}

\subsection{Identification of grand minima and maxima}

Using the criteria defined in section~\ref{sec:analysesIdent}, we identified grand minimum and maximum events in the experiments, in which random fluctuations in the BL-mechanism are used (Table~\ref{tab:RNBL}). For the correlation time-scale of $\sim16$ days ($\tau_{\rm c}=4$\,$\Omega_{0}^{-1}$), only the fluctuations with the amplitude of 1.0 show grand minima and maxima, while for the correlation time-scales of $\sim64$ days ($\tau_{\rm c}=16$\,$\Omega_{0}^{-1}$) fluctuations with the amplitudes between 0.5 and 1.0 show grand minima and maxima (Table~\ref{tab:RNBL}).

We show an example of the identified grand minima and maxima generated by applying random fluctuations in the BL-mechanism in experiment 38 (Table~\ref{tab:RNBL}) in the top panel of Figure~\ref{fig:E38MinMax}, while the bottom panel represents the butterfly diagram of $B_{\phi}$ at 0.71$R_{\odot}$ for the minimum event identified between $\sim5890$--5940. Similar to the butterfly diagram for the minimum period in experiment 18, the toroidal field strength at 0.71$R_{\odot}$ is considerably smaller compared to before and after the minimum.

\subsection{Waiting Time Distributions}

\begin{figure*}
\begin{center}
{\includegraphics[width=5.2in]{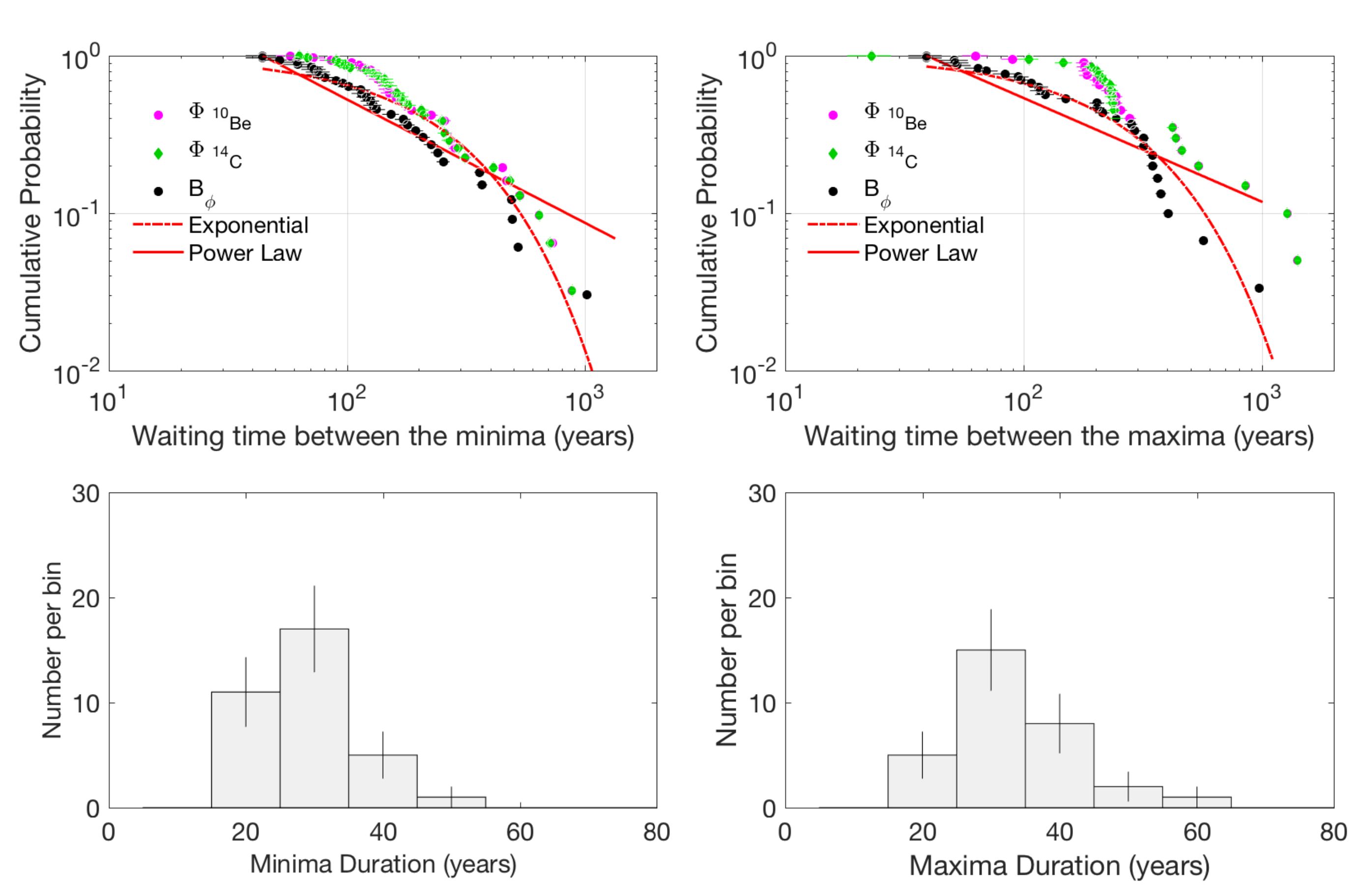}}
\caption{The top panel shows the WTD of grand minima and maxima from experiment 35 together with the observed data points from $^{10}$Be and $^{14}$C based solar modulation potentials from \citet{2015A&A...577A..20I}. The bottom panel shows the distribution of the durations of grand minima and maxima.} 
\label{fig:E35WTDDur}
\end{center}
\end{figure*}

Performed WTD analyses, where we investigate whether they show an exponential or a power-law distribution, show that almost all of the WTDs of grand minima and maxima from experiments are best represented with an exponential distribution, except for the experiment 35's WTD of grand minima (Table~\ref{tab:gminmaxBL}), which is shown in Figure~\ref{fig:E35WTDDur} together with the WTDs of grand solar minima and maxima from the $^{10}$Be and $^{14}$C based solar modulation potentials from \citet{2015A&A...577A..20I}.

The results revealed that the distributions of durations of grand minima and maxima are log-normal distributions for all experiments, and the average durations are clustered around $\sim28$ years for grand minima and $\sim34$ year for grand maxima (Table~\ref{tab:gminmaxBL}).

\begin{figure*}[htb!]
\begin{center}
{\includegraphics[width=5.5in]{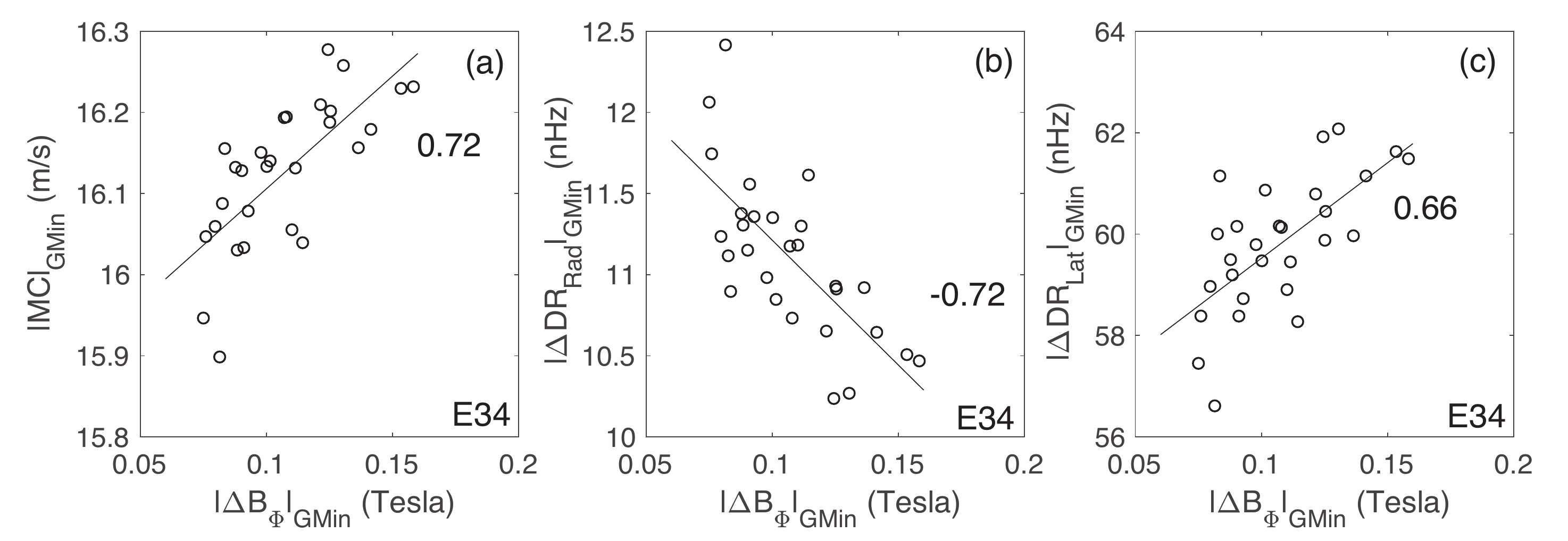}}
\caption{The panel (a) shows the correlation between the amplitudes of surface meridional circulation at 45$^{\circ}$ latitude and the amplitudes of grand minima for experiment 34 (R=0.72). The panel (b) shows the correlation between the amplitudes of difference in rotation rates between 0.70$R_{\odot}$ and 0.985$R_{\odot}$, and at 45$^{\circ}$ latitude ($\Delta{\rm DR_{Rad}}$) and the amplitudes of grand minima (R=-0.72), while the panel (c) shows the correlation between the amplitudes of difference in rotation rates between 0$^{\circ}$ and 60$^{\circ}$ latitudes and at 0.985$R_{\odot}$ ($\Delta{\rm DR_{Lat}}$) and the amplitudes of grand minima for experiment 34 (R=0.66).}
\label{fig:DRMCRB}
\end{center}
\end{figure*}

\subsection{Meridional Circulation and Differential Rotation}

Similar to the results obtained for the random fluctuations in the $\Lambda$-mechanism, the results from the two sample KS tests revealed that the MC during the grand minimum periods are statistically different and faster than those during the grand maximum periods, where it is slower, at 99\% significance level. The $k_{\rm MC}$ values calculated for the experiments show that the greatest difference in the MC during grand minima and maxima is observed for experiment 38, while experiment 33 has the smallest difference in MC values. The results also show that the $k_{\rm MC}$ values are relatively larger in comparison to those calculated for the fluctuations in the $\Lambda$-mechanism (Table~\ref{tab:2KSRLam}).

\begin{table}[ht]
\centering
\caption{The test statistics values for the two sample KS test for meridional circulation at 45$^{\circ}$ latitude ($k_{\rm MC}$), and the difference in rotation rates between 0.70$R_{\odot}$ and 0.985$R_{\odot}$, and at 45$^{\circ}$ latitude ($k_{\Delta {\rm DR_{Rad}}}$) and the difference in rotation rates between 0$^{\circ}$ and 60$^{\circ}$ latitudes and at 0.985$R_{\odot}$ ($k_{\Delta{\rm DR_{Lat}}}$). Larger $k$ values represent larger differences.}
\begin{tabular}{lccc}
\hline \hline
E	&	$k_{\rm MC}$	&	$k_{\Delta{\rm DR_{Rad}}}$		&$k_{\Delta{\rm DR_{Lat}}}$		\\ \hline
28	&	0.63			&	0.66							& 0.41	\\
33	&	0.56			&	0.59							& 0.40	\\
34	&	0.62			&	0.60							& 0.41	\\
35	&	0.57			&	0.56							& 0.39	\\
36	&	0.62			&	0.65							& 0.43	\\
37	&	0.67			&	0.66							& 0.48	\\
38	&	0.69			&	0.64							& 0.45	\\
\hline
\end{tabular}
\label{tab:2KSRBL}
\end{table}

For the $\Delta{\rm DR_{Rad}}$ values during the grand minima and maxima, the two sample KS tests showed that they are statistically different at 99\% significance level and the $\Delta{\rm DR_{Rad}}$ values during grand minimum are smaller than those during the maximum periods, which are larger. The $k_{\Delta{\rm DR_{Rad}}}$ values for the experiments show that experiment 35 show the smallest $\Delta{\rm DR_{Rad}}$ between grand minimum and maximum periods, while experiments 28 and 37 has the greatest difference. The results also show that the $k_{\Delta{\rm DR_{Rad}}}$ values are greater than $k_{\Delta{\rm DR_{Rad}}}$ values calculated for the random fluctuations in the $\Lambda$-mechanism (Table~\ref{tab:2KSRLam}). 

Similar to $\Delta{\rm DR_{Rad}}$ values, the difference between the rotation values calculated at 0$^{\circ}$ and 60$^{\circ}$ latitudes and at 0.985$R_{\odot}$ are also statistically different at 99\% significance level. The $\Delta{\rm DR_{Lat}}$ values are larger during grand minima, while they are smaller during grand maxima. The $k_{\Delta{\rm DR_{Rad}}}$ values show that the greatest difference in $\Delta{\rm DR_{Lat}}$ is observed in experiment 37, while experiment 35 shows the smallest (Table~\ref{tab:2KSRBL}). 

Figure~\ref{fig:DRMCRB}a, b, and c show the relationship between the amplitudes of surface meridional circulation at 45$^{\circ}$ latitude and the amplitudes of grand minima (R=0.72), between the amplitudes of difference in rotation rates between 0.70$R_{\odot}$ and 0.985$R_{\odot}$, and at 45$^{\circ}$ latitude ($\Delta{\rm DR_{Rad}}$) and the amplitudes of grand minima (R=-0.72), and between the amplitudes of difference in rotation rates between 0$^{\circ}$ and 60$^{\circ}$ latitudes and at 0.985$R_{\odot}$ ($\Delta{\rm DR_{Lat}}$) and the amplitudes of grand minima for experiment 34 (R=0.66).

\subsection{Role of the Lorentz force on the meridional circulation and difference in rotation rates}

The variations in the meridional circulation at 45$^{\circ}$ latitude and the difference in rotation rates between 0.70$R_{\odot}$ and 0.985$R_{\odot}$ at 45$^{\circ}$ latitude, and the difference in rotation rates between 0$^{\circ}$ and 60$^{\circ}$ latitudes and at 0.985$R_{\odot}$ during grand minima and maxima obtained from the simulations where random fluctuations are introduced into the $\Lambda$-and-BL-mechanisms are similar. For the both cases under consideration, the results show that the meridional circulation seems to be faster and the difference in rotation rate is small during grand minima. This might suggest that the variations in the meridional flow and the rotation rates are likely dominated by the Lorentz forces of the dynamo field. To investigate the possible effects of the Lorentz force on the meridional circulation and the differential rotation, and also on the nature and occurrence characteristic of grand minima and maxima, we ran three additional simulations without the Lorentz force included, using the parameters for experiments 18 and 38 (hereafter E18-NL and E38-NL). Because we disabled the Lorentz force feedback, which originally acts as a saturation mechanism \citep[see][for details]{2006ApJ...647..662R} for the dynamo, we enabled a small amplitude $\alpha$-quenching to replace the role of the Lorentz force. This also led us use the $\alpha$-coefficient with an amplitude of 0.125 m\,s$^{-1}$, instead of 0.4 m\,s$^{-1}$ used in the simulations what include the Lorentz force. Additionally, we ran a hydrodynamical simulation without any magnetic field generation that has random fluctuations (hereafter HD-RL) as experiment 18. 

\begin{figure*}[htb!]
\begin{center}
{\includegraphics[width=\textwidth]{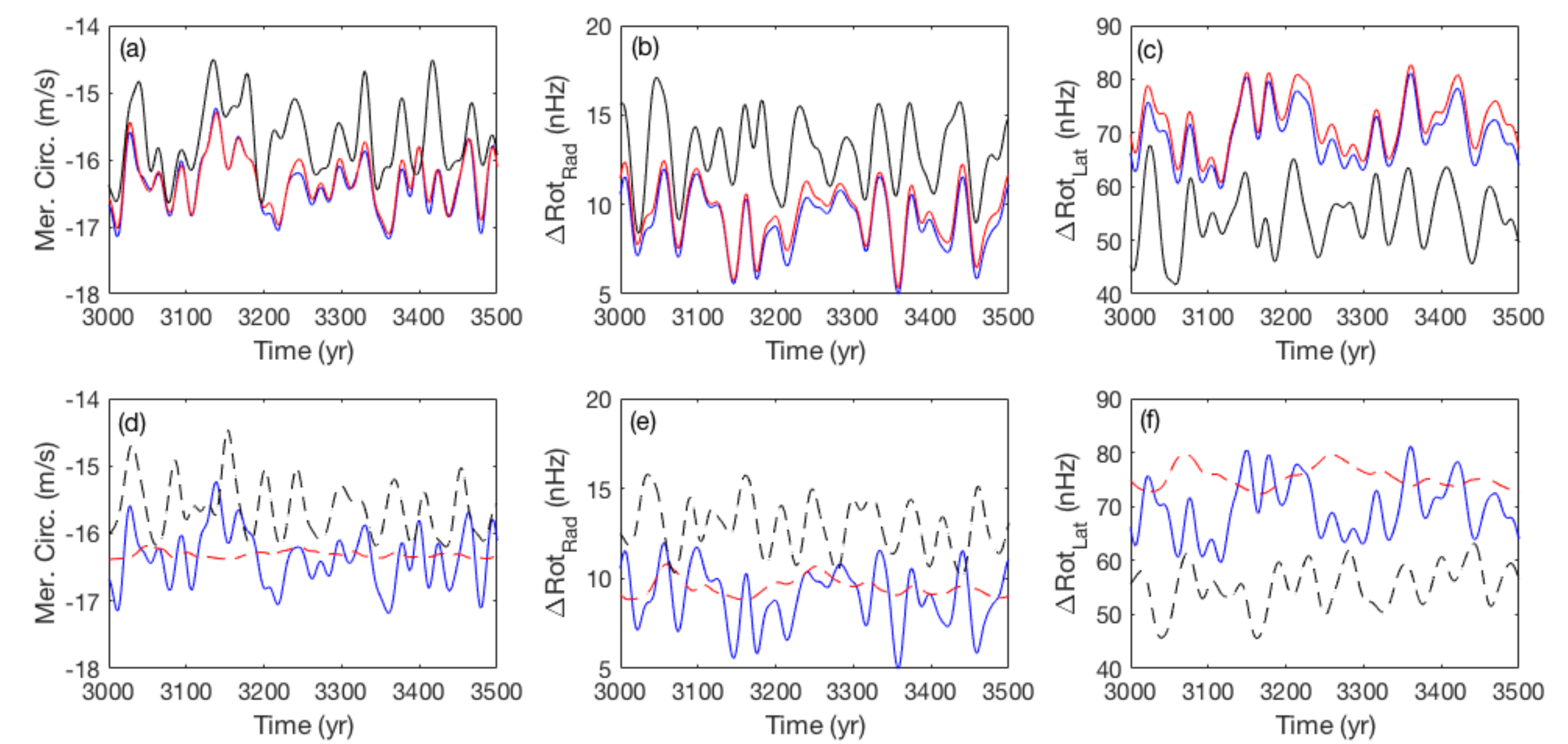}}
\caption{Panel (a) shows the surface meridional circulation at 45$^{\circ}$ latitude for experiment 18 (solid black), E18-NL (solid red), and HD-RL (solid blue), while panel (d) shows the same variations for experiment 38 (dashed black), E18-NL (dashed red), and HD-RL (solid blue). Panel (b) shows the difference in rotation rates between 0.70$R_{\odot}$ and 0.985$R_{\odot}$, and at 45$^{\circ}$ latitude ($\Delta{\rm DR_{Rad}}$) for experiment 18 (solid black), E18-NL (solid red), and HD-RL (solid blue), while panel (e) shows the same for experiment 38 (dashed black), E18-NL (dashed red), and HD-RL (solid blue). Panel (c) shows the difference in rotation rates between 0$^{\circ}$ and 60$^{\circ}$ latitudes and at 0.985$R_{\odot}$ ($\Delta{\rm DR_{Lat}}$) for experiment 18 (solid black), E18-NL (solid red), and HD-RL (solid blue), while panel (f) shows the same for experiment 38 (dashed black), E18-NL (dashed red), and HD-RL (solid blue). The negative sign in the y-axis of  panels (a) and (d) shows the pole-ward flow on top of the domain at 0.985$R_{\odot}$.}
\label{fig:RoleLorentz}
\end{center}
\end{figure*}

The results show that the Lorentz force causes the variations in the meridional circulation and in the difference in rotation rates to have larger amplitudes (black solid and dashed curves in Figure~\ref{fig:RoleLorentz}), and greater averages for experiment 18 compared to E18-NL (solid black and red curves in the top panels of Figure~\ref{fig:RoleLorentz}, respectively). For E38-NL, the variations in the meridional circulation and the rotations rates have considerably small amplitudes (dashed red curves in the bottom panels of Figure~\ref{fig:RoleLorentz}), which might be caused by the ohmic heat in the entropy equation \citep{2005ApJ...622.1320R}, while for experiment 38 the amplitude of variations in both the meridional circulation and the differential rotation are larger (dashed black curves in the bottom panels of Figure~\ref{fig:RoleLorentz}). As for the HD-RL run, the fluctuations in the meridional circulation and the differential rotation, which are introduced by random fluctuations in the $\Lambda$-mechanism, show amplitudes similar to that E18-NL (blue curves in the top panels of Figure~\ref{fig:RoleLorentz}). These amplitudes are also considerably larger than those in E38-NL, but smaller than those observed in experiments 18 and 38.

We also investigated the influence of the Lorentz force on the nature and occurrence statistics of grand minima and maxima in E18-NL and E38-NL. The results showed that there are 28 grand maxima and 17 grand minima identified in E18-NL, while there are 25 grand maxima and 9 grand minima found in E38-NL. The number of grand minima and maxima identified in E18-NL and in E38-NL are smaller than those found for experiments 18 and 38, where the Lorentz force is enabled (see Tables~\ref{tab:RNLambda} and~\ref{tab:RNBL}, respectively). The WTD analyses show that the distributions of the grand minima and maxima identified in E18-NL are better represented by exponential distributions. In contrast to experiment 18, disabling the Lorentz force caused the distribution of the durations of grand minima and maxima to shift to larger values that are $\sim70$ and $\sim60$ years, respectively. For E38-NL, the WTD analyses show that the distribution of grand maxima is better represented with an exponential distribution, while a power-law represented the distribution of the grand minima better. It must be emphasised that there are only 8 waiting times (9 grand minima identified in this simulation) and the result might not be very representative. Similar to the results from E18-NL, the distributions of the durations of grand maxima and grand minima identified in E38-NL shifted towards larger values, which are $\sim60$ and $\sim97$ years, respectively.

\section{Discussion and Conclusions} \label{sec:discconc}

The results from the WTD analyses of the identified grand minimum and maximum periods in our experiments, where random fluctuations in the $\Lambda$-mechanism was used to generate disturbances in the large-scale magnetic field of the dynamo, revealed that the majority of the distributions of the waiting times between grand minima and grand maxima are better represented by exponential distributions. This means that, as a physical mechanism, the random fluctuations in the $\Lambda$-mechanism are translated into the occurrences of grand minima and maxima as memoryless-processes as expected. A few exceptional cases, such as the WTDs of grand minima and maxima for experiment 9 and the WTDs of grand maxima for experiment 11, are better represented by power-law distributions. To test whether the results are dependent on the length of the time series, we ran experiment 9 for 24208 years and performed WTD analyses for the identified grand minima and maxima. The results from the two sample KS test show that the distributions of the WTDs of grand minima ($p_{\rm pow}$=0.001, $p_{\rm exp}$=0.02) and maxima ($p_{\rm pow}$=0.008, $p_{\rm exp}$=0.04) are better represented by exponential distributions for longer simulations, instead of power-law for shorter ones. These results might indicate the importance of the length of the time series.

Similar to the results obtained for the fluctuations in the $\Lambda$-mechanism, we found that the WTDs of grand minimum and maximum periods identified in the experiments, where we used random fluctuations in the BL-mechanism, are better represented by exponential distributions. For experiments 35 and 37, the WTDs of grand minima are better represented by power-laws, while for the WTD of grand minima of experiment 38, the result is inconclusive. 

We also tested whether having longer correlation time-scales of 256 days ($\tau_{\rm c}=64$\,$\Omega_{0}^{-1}$$\approx256$ days) would have an impact on the results found for the correlation time-scales of 16 and 64 days. To achieve this, we used the parameters used in experiments 3, 9 and 15 for the random fluctuations in the $\Lambda$-mechanism (Table~\ref{tab:RNLambda}) and those in experiments 33, 36, and 38 for the random fluctuations in the BL-mechanism (Table~\ref{tab:RNBL}). The only difference in these simulations was the correlation time-scale, which is used as $\tau_{\rm c}=64$\,$\Omega_{0}^{-1}$$\approx256$ days. The results show that having longer correlation time-scales has a small impact on the amplitudes of the grand minimum and maximum periods, however the statistical outputs remain the same.

Previously, \citet{2008SoPh..250..221M} and \citet{2009SoPh..254..345U} generated grand solar minimum-like periods by applying random fluctuations in the $\alpha$-coefficient of the Parker migratory dynamo model they used, and performed WTD analyses for the grand minimum periods they identified in their simulations. The results from the two studies showed that random fluctuations in the $\alpha$-coefficient is capable of generating grand solar minimum-like periods and the WTDs of these events are better represented by exponential distributions, meaning that these periods occur as a result of a Poisson process. More recently, \citet{2017arXiv170510746C} claimed that the decadal to millennial variations in solar activity can be generated with a weakly non-linear and noisy limit cycle, under the assumptions that the 11-year solar cycle had continued during the Maunder Minimum, the linear growth rate, which implies a recovery from the Maunder Minimum, is in the order of $1/50$~yr$^{-1}$, and the noise levels are around 35\%. The resulting WTDs are therefore of exponential shape, i.e. of stochastic origin. In their work, \citet{2008SoPh..250..221M} pointed out that some of their simulations initially showed power-law distributions, all of which changed to exponential distributions provided longer time series, which are consistent with our results for experiment 9. Our findings, together with of \citet{2008SoPh..250..221M} and \citet{2009SoPh..254..345U}, seems to contradict the results obtained from cosmogenic isotope records ($^{10}$Be, $^{14}$C), which show that the WTDs of grand solar minima and maxima are better represented by a power-law, indicating that there is a memory effect in the occurrences of these high-and low-activity periods \citep{2007A&A...471..301U,2015A&A...577A..20I}. However, \citet{2009SoPh..254..345U} suggested that the probability of finding a power-law distribution from a subset of purely exponentially distributed data increases with decreasing number of events \citep[see Figure 3 in][]{2009SoPh..254..345U}. This means that the $^{10}$Be, $^{14}$C records used to investigate the past variations in the solar activity levels are not long enough to study the occurrence characteristics of grand solar minimum and maximum periods. It should be noted though that \citet{2009SoPh..254..345U} used the logarithmic least square method to fit their exponential and power-law distributions, which is criticised by \citet{Clauset2009} because this method is prone to generate systematic errors and inaccurate estimates of power-law distribution parameters. The results should therefore be evaluated with caution.

The cosmogenic nuclides, $^{10}$Be and $^{14}$C, revealed that the solar cycles continued during grand minimum periods like the Maunder Minimum \citep{2012GeoRL..3919102O,2013SSRv..176...59M,2014SoPh..289.4377I}. Using $^{14}$C records, \citet{2004SoPh..224..317M} claimed that the lengths of the solar cycles during the Maunder Minimum tend to be longer, while using a $^{10}$Be record, \citet{1998SoPh..181..237B} suggested that the length of the solar cycles during the Maunder Minimum was around 11-years. We check our simulations, which were not smoothed, for variations in the solar cycle lengths. The periods range from $\sim8$ years to $\sim15$ years, which is in agreement with the observed variations in the solar cycle length. At first glance, we could not find any systematic change during grand minimum and maximum periods. However, a thorough investigation of periods and cycle length variations in the simulations are left for a follow-up paper, since observational results in this are still uncertain.

The durations of the grand minima and maxima identified in our experiments, where the random fluctuations are introduced to the $\Lambda$-mechanism are clustered around 33 and 36 years, respectively, while for the fluctuations in the BL-mechanism lead grand minima and maxima to have durations clustered around 28 and 34 years, respectively. It should be noted that even though the correlation time-scales of the fluctuations in the $\Lambda$-and-BL mechanisms used in our study are in the order of days, the resulting durations are in order of years. The results from the cosmogenic nuclides, $^{10}$Be and $^{14}$C, show that the durations of grand minima and maxima events are around 65 and 70 years, respectively, with upper limits reaching around 175 years for grand minimum and 100 years for grand maximum periods \citep{2007A&A...471..301U,2015A&A...577A..20I}. Using the results from their BL-type flux transport dynamo simulations, \citet{2013RAA....13.1339K} suggested that there is a positive relationship between the coherence time of the fluctuations in the meridional circulation and the BL-mechanism, and the duration and number of grand minimum events. However, it must be noted that the coherence times used in their study varies between 10--50 years, while we used 16 and 64 days.

The meridional circulation during the grand minimum and maximum periods in our experiments, independent of the physical mechanism causing these periods (random fluctuations in the $\Lambda$- or BL-mechanisms), shows statistically significant differences. The results of the two sample KS test show that the meridional circulation on the domain surface at 45$^{\circ}$ latitude is slower during grand maximum periods than that during grand minima. Our results contradict those derived using a diffusion-dominated flux transport dynamo by \citet{2010ApJ...724.1021K}, who suggested that the meridional circulation starts to decrease before entering the Maunder Minimum and it recovers towards the termination of the period after staying lower for a while. However, it must be noted that the magnetic diffusivity for diffusion-dominated model used in \citet{2010ApJ...724.1021K} is $\sim$10$^{12}$--10$^{13}$ cm$^{2}$\,s$^{-1}$ in the whole convection zone. In our model, the magnetic diffusivity decreases gradually with depth starting from 10$^{12}$ cm$^{2}$\,s$^{-1}$ on top of the domain (R=0.985$R_{\odot}$) to 10$^{10}$ cm$^{2}$\,s$^{-1}$ in the bottom of the domain ($R=0.65R_{\odot}$) \citep{2006ApJ...647..662R}. In addition, \citet{2010ApJ...724.1021K} reported that for their advection-dominated model, the magnetic diffusivity of which is $\sim$10$^{10}$--10$^{11}$ cm$^{2}$\,s$^{-1}$ in the whole convection zone, does not show any relationship between the meridional flow speed and the strength of the magnetic field. The reason for this difference is that for lower magnetic diffusivities, slower meridional flow speeds lead to stronger toroidal field in the solar tachocline, since the pre-existing poloidal field has more time to build up. Hence, slower meridional circulation causes generation of a stronger toroidal field. For higher magnetic diffusivities, slower meridional flow speeds means that during its transportation throughout the convection zone, there will be more time for the diffusive decay of the poloidal field, which in turn generates weaker toroidal field in the solar tachocline. 

The results from the performed two sample KS test on the difference between the differential rotation rates at 0.70$R_{\odot}$ and at 0.985$R_{\odot}$ show that the difference is larger during the grand maximum periods than those calculated during the grand minimum periods. On the other hand, the difference between the rotation rates at 0$^{\circ}$ and 60$^{\circ}$ latitudes and at 0.985$R_{\odot}$ is found to be larger during grand minima at 99\% significance level. Sunspot observations spanning from 1666 to 1719 indicated that the solar rotation became more differential during the Maunder Minimum, where the rotation rate of the Sun at the equator was longer around 27.5 days \citep{1993A&A...276..549R}, whereas the differential rotation rates after the Maunder Minimum, between 1749--1799, were similar to the value of the present Sun \citep{2012A&A...543A...7A}.

The correlations between the amplitudes of surface meridional circulation at 45$^{\circ}$ latitude and the amplitudes of grand minima are found to be lower in simulations where random fluctuations are introduced in the $\Lambda$-mechanism in comparison to those in random fluctuations in Babcock-Leighton mechanism, where they are considerably higher. The same situation is also observed for the correlations between the amplitudes of difference in rotation rates between 0.70$R_{\odot}$ and 0.985$R_{\odot}$, and at 45$^{\circ}$ latitude ($\Delta{\rm DR_{Rad}}$) and the amplitudes of grand minima, and between the amplitudes of difference in rotation rates between 0$^{\circ}$ and 60$^{\circ}$ latitudes and at 0.985$R_{\odot}$ ($\Delta{\rm DR_{Lat}}$).

The impact of the Lorentz force on the meridional circulation and the rotation rates showed that when the Lorentz force is disabled, the fluctuations in these flows have smaller amplitudes, meaning that the Lorentz force enhances the variability in the flow field. The results also showed that the durations of grand minima and maxima are mainly controlled by the Lorentz force, where they are almost two folds longer in its absence.

It was suggested that a recovery mechanism is needed after a grand minimum period in simulations from Babcock-Leighton type dynamos, where there are no bipolar active region to support generation of a poloidal field \citep{2012PhRvL.109q1103C,2014ApJ...789....5H}. In our dynamo models, there is not a threshold value for the Babcock-Leighton $\alpha$-effect, which leads the dynamo to recover from low activity periods like grand minima. This approach is based on the fact that sunspots are the upper end of the spectrum of flux emergence and even if they disappear, there are still ephemeral regions that will still obey the Hale's polarity law to a small degree \citep{Priest2014}.

In conclusion, our study showed that a Babcock-Leighton-type flux transport dynamo with random fluctuations in the $\Lambda$- and BL-mechanisms are capable of generating grand solar minima and maxima-like periods. However, the WTDs of these events do not agree with the results drawn from the cosmogenic isotope records. Further, the average durations of grand minimum and maximum periods identified in our simulations, together with their upper limits, are in disagreement with those from the cosmogenic nuclides. Our results showed that the meridional flow speed is higher during grand minima, which may be linked to the low magnetic diffusivity. Under this condition, the poloidal field that has less time to build up a stronger toroidal field when the meridional flow is faster. The similar behaviour of the meridional circulation and the difference in rotation rates during grand minimum and maximum periods observed in the simulations with fluctuations in the $\Lambda$-mechanism and the BL-mechanism might suggest that the variations in these flows are mainly caused by the Lorentz forces of the dynamo field, and these forces could be the possible cause of the observed grand minima and maxima. The variability of the solar cycle is more likely caused by a non-linear dynamical system rather than a purely stochastically perturbed one.

This study can be regarded as a reference study for a future research. We plan to include the both hemispheres as well as to use different diffusivities in a distributed $\alpha$-dynamo to better investigate the role of the Lorentz forces and the variations in the meridional flow and the differential rotation on the strength of the toroidal field and parity change.

\acknowledgments

FI acknowledges the Carlsberg Foundation (CF15-0648). The National Center for Atmospheric Research is sponsored by the National Science Foundation. We thank Bidya Karak for his useful comments.




\begin{thebibliography}{}

\bibitem[Arlt \& Fr{\"o}hlich(2012)]{2012A&A...543A...7A} Arlt, R., \& Fr{\"o}hlich, H.-E.\ 2012, \aap, 543, A7

\bibitem[Babcock(1961)]{1961ApJ...133..572B} Babcock, H.~W.\ 1961, \apj, 133, 572

\bibitem[Beer et al.(1990)]{1990Natur.347..164B} Beer, J., Blinov, A., Bonani, G., Hofmann, H.~J., \& Finkel, R.~C.\ 1990, \nat, 347, 164 

\bibitem[Beer et al.(1998)]{1998SoPh..181..237B} Beer, J., Tobias, S., \& Weiss, N.\ 1998, \solphys, 181, 237 

\bibitem[Belucz \& Dikpati(2013)]{2013ApJ...779....4B} Belucz, B., \& Dikpati, M.\ 2013, \apj, 779, 4

\bibitem[Brandenburg \& Subramanian(2005)]{2005PhR...417....1B} Brandenburg, A., \& Subramanian, K.\ 2005, \physrep, 417, 1 

\bibitem[Braj{\v s}a et al.(2006)]{2006SoPh..237..365B} Braj{\v s}a, R., Ru{\v z}djak, D., \& W{\"o}hl, H.\ 2006, \solphys, 237, 365

\bibitem[Bushby(2006)]{2006MNRAS.371..772B} Bushby, P.~J.\ 2006, \mnras, 371, 772

\bibitem[Cameron \& Sch{\"u}ssler(2010)]{2010ApJ...720.1030C} Cameron, R.~H., \& Sch{\"u}ssler, M.\ 2010, \apj, 720, 1030

\bibitem[Cameron \& Sch{\"u}ssler(2015)]{2015Sci...347.1333C} Cameron, R., \& Sch{\"u}ssler, M.\ 2015, Science, 347, 1333

\bibitem[Cameron \& Sch{\"u}ssler(2017)]{2017arXiv170510746C} Cameron, R., \& Sch{\"u}ssler, M.\ 2017, arXiv:1705.10746 

\bibitem[Charbonneau(2014)]{2014ARA&A..52..251C} Charbonneau, P.\ 2014, \araa, 52, 251

\bibitem[Choudhuri \& Karak(2009)]{2009RAA.....9..953C} Choudhuri, A.~R., \& Karak, B.~B.\ 2009, Research in Astronomy and Astrophysics, 9, 953

\bibitem[Choudhuri \& Karak(2012)]{2012PhRvL.109q1103C} Choudhuri, A.~R., \& Karak, B.~B.\ 2012, Physical Review Letters, 109, 171103

\bibitem[Clauset et al.(2009)]{Clauset2009} Clauset, A., Rohilla Shalizi, C., \& Newman, M.~E.~J. 2009, Society for Industrial and Applied Mathematics Rev., 51, 4. 

\bibitem[Corsaro et al.(2013)]{2013MNRAS.430.2313C} Corsaro, E., Fr{\"o}hlich, H.-E., Bonanno, A., et al.\ 2013, \mnras, 430, 2313

\bibitem[Dikpati \& Charbonneau(1999)]{1999ApJ...518..508D} Dikpati, M., \& Charbonneau, P.\ 1999, \apj, 518, 508

\bibitem[Dikpati \& Gilman(2001)]{2001ApJ...559..428D} Dikpati, M., \& Gilman, P.~A.\ 2001, \apj, 559, 428

\bibitem[Dunai(2010)]{Dunai2010} Dunai, T. J. 2010, Cosmogenic Nuclides: Principles, Concepts and Applications in the Earth Surface Sciences, Cambridge University Press, Cambridge, 2

\bibitem[Gonz{\'a}lez Hern{\'a}ndez et al.(2008)]{2008SoPh..252..235G} Gonz{\'a}lez Hern{\'a}ndez, I., Kholikov, S., Hill, F., Howe, R., \& Komm, R.\ 2008, \solphys, 252, 235

\bibitem[Guerriero (2012)]{Guerriero2012} Guerriero, V., 2012, Journal of Modern Mathematics Frontier, 1, 21

\bibitem[Hathaway \& Rightmire(2010)]{2010Sci...327.1350H} Hathaway, D.~H., \& Rightmire, L.\ 2010, Science, 327, 1350

\bibitem[Hazra et al.(2014)]{2014ApJ...789....5H} Hazra, S., Passos, D., \& Nandy, D.\ 2014, \apj, 789, 5

\bibitem[Howard(1996)]{1996ARA&A..34...75H} Howard, R.~F.\ 1996, \araa, 34, 75

\bibitem[Howard \& Labonte(1980)]{1980ApJ...239L..33H} Howard, R., \& Labonte, B.~J.\ 1980, \apjl, 239, L33

\bibitem[Howe et al.(2000)]{2000Sci...287.2456H} Howe, R., Christensen-Dalsgaard, J., Hill, F., et al.\ 2000, Science, 287, 2456

\bibitem[Howe(2009)]{2009LRSP....6....1H} Howe, R.\ 2009, Living Reviews in Solar Physics, 6, 1 

\bibitem[Inceoglu et al.(2014)]{2014SoPh..289.4377I} Inceoglu, F., Knudsen, M.~F., Karoff, C., \& Olsen, J.\ 2014, \solphys, 289, 4377

\bibitem[Inceoglu et al.(2015)]{2015A&A...577A..20I} Inceoglu, F., Simoniello, R., Knudsen, M.~F., et al.\ 2015, \aap, 577, A20

\bibitem[Inceoglu et al.(2016)]{2016SoPh..291..303I} Inceoglu, F., Simoniello, R., Knudsen, M.~F., et al.\ 2016, \solphys, 291, 303 

\bibitem[Karak(2010)]{2010ApJ...724.1021K} Karak, B.~B.\ 2010, \apj, 724, 1021

\bibitem[Karak \& Choudhuri(2013)]{2013RAA....13.1339K} Karak, B.~B., \& Choudhuri, A.~R.\ 2013, Research in Astronomy and Astrophysics, 13, 1339-1357 

\bibitem[Kitchatinov et al.(1994)]{1994A&A...292..125K} Kitchatinov, L.~L., Ruediger, G., \& Kueker, M.\ 1994, \aap, 292, 125

\bibitem[Kitchatinov \& R{\"u}diger(2005)]{2005AN....326..379K} Kitchatinov, L.~L., \& R{\"u}diger, G.\ 2005, Astronomische Nachrichten, 326, 379 

\bibitem[Knudsen et al.(2009)]{2009GeoRL..3616701K} Knudsen, M.~F., Riisager, P., Jacobsen, B.~H., et al.\ 2009, \grl, 36, L16701

\bibitem[Komm et al.(2015)]{2015SoPh..290.3113K} Komm, R., Gonz{\'a}lez Hern{\'a}ndez, I., Howe, R., \& Hill, F.\ 2015, \solphys, 290, 3113

\bibitem[K{\"u}ker et al.(1999)]{1999A&A...343..977K} K{\"u}ker, M., Arlt, R., \& R{\"u}diger, G.\ 1999, \aap, 343, 977 

\bibitem[Lean et al.(2002)]{2002GeoRL..29.2224L} Lean, J.~L., Wang, Y.-M., \& Sheeley, N.~R.\ 2002, \grl, 29, 2224

\bibitem[Leighton(1964)]{1964ApJ...140.1547L} Leighton, R.~B.\ 1964, \apj, 140, 1547

\bibitem[Lepreti et al.(2001)]{2001ApJ...555L.133L} Lepreti, F., Carbone, V., \& Veltri, P. 2001, \apjl, 555, L133

\bibitem[Malkus \& Proctor(1975)]{1975JFM....67..417M} Malkus, W.~V.~R., \& Proctor, M.~R.~E.\ 1975, Journal of Fluid Mechanics, 67, 417 

\bibitem[McCracken et al.(2013)]{2013SSRv..176...59M} McCracken, K., Beer, J., Steinhilber, F., \& Abreu, J.\ 2013, \ssr, 176, 59.

\bibitem[Miyahara et al.(2004)]{2004SoPh..224..317M} Miyahara, H., Masuda, K., Muraki, Y., et al.\ 2004, \solphys, 224, 317

\bibitem[Moss et al.(2008)]{2008SoPh..250..221M} Moss, D., Sokoloff, D., Usoskin, I., \& Tutubalin, V.\ 2008, \solphys, 250, 221

\bibitem[Nandy \& Choudhuri(2001)]{2001ApJ...551..576N} Nandy, D., \& Choudhuri, A.~R.\ 2001, \apj, 551, 576

\bibitem[Nandy \& Choudhuri(2002)]{2002Sci...296.1671N} Nandy, D., \& Choudhuri, A.~R.\ 2002, Science, 296, 1671

\bibitem[Olemskoy \& Kitchatinov(2013)]{2013ApJ...777...71O} Olemskoy, S.~V., \& Kitchatinov, L.~L.\ 2013, \apj, 777, 71

\bibitem[Owens et al.(2012)]{2012GeoRL..3919102O} Owens, M.~J., Usoskin,I., \& Lockwood, M. 2012, \grl, 39, 19102

\bibitem[Parker (1955a)]{1955ApJ...122..293P} Parker, E.~N., 1955a, \apj, 122, 293

\bibitem[Parker (1955b)]{1955ApJ...121..491P} Parker, E.~N., 1955b, \apj, 121, 491

\bibitem[Passos et al.(2014)]{2014A&A...563A..18P} Passos, D., Nandy, D., Hazra, S., \& Lopes, I.\ 2014, \aap, 563, A18

\bibitem[Pipin(1999)]{1999A&A...346..295P} Pipin, V.~V.\ 1999, \aap, 346, 295

\bibitem[Potgieter(2013)]{2013LRSP...10....3P} Potgieter, M.~S.\ 2013, Living Reviews in Solar Physics, 10, 3 

\bibitem[Priest(2014)]{Priest2014} Priest, E.\ 2014, Cambridge University Press, 25-30.

\bibitem[Rempel(2005a)]{2005ApJ...622.1320R} Rempel, M.\ 2005, \apj, 622, 1320 

\bibitem[Rempel(2005b)]{2005ApJ...631.1286R} Rempel, M.\ 2005, \apj, 631, 1286

\bibitem[Rempel(2006)]{2006ApJ...647..662R} Rempel, M.\ 2006, \apj, 647, 662

\bibitem[Rempel(2007)]{2007AN....328.1096R} Rempel, M.\ 2007, Astronomische Nachrichten, 328, 1096

\bibitem[Ribes \& Nesme-Ribes(1993)]{1993A&A...276..549R} Ribes, J.~C., \& Nesme-Ribes, E.\ 1993, \aap, 276, 549

\bibitem[Schou et al.(2002)]{2002ApJ...567.1234S} Schou, J., Howe, R., Basu, S., et al.\ 2002, \apj, 567, 1234 

\bibitem[Senthamizh Pavai et al.(2016)]{2016AdSpR..58.1468S} Senthamizh Pavai, V., Arlt, R., Diercke, A., Denker, C., \& Vaquero, J.~M.\ 2016, Advances in Space Research, 58, 1468

\bibitem[Steinhilber et al.(2012)]{2012PNAS..109.5967S} Steinhilber, F., Abreu, J.~A., Beer, J., et al.\ 2012, Proceedings of the National Academy of Science, 109, 5967

\bibitem[Thompson et al.(2003)]{2003ARA&A..41..599T} Thompson, M.~J., Christensen-Dalsgaard, J., Miesch, M.~S., \& Toomre, J.\ 2003, \araa, 41, 599

\bibitem[Tobias(1996)]{1996A&A...307L..21T} Tobias, S.~M.\ 1996, \aap, 307, L21

\bibitem[Tobias(1997)]{1997A&A...322.1007T} Tobias, S.~M.: 1997, \aap, 322, 1007

\bibitem[Usoskin(2013)]{2013LRSP...10....1U} Usoskin, I.~G.: 2013, Liv. Rev. in Solar Phys., 10, 1

\bibitem[Usoskin et al.(2009)]{2009SoPh..254..345U} Usoskin, I.~G., Sokoloff, D., \& Moss, D.\ 2009, \solphys, 254, 345

\bibitem[Usoskin et al.(2007)]{2007A&A...471..301U} Usoskin, I.~G., Solanki, S.~K., \& Kovaltsov, G.~A. 2007, \aap, 471, 301

\bibitem[Virkar \& Clauset(2014)]{VirkarandClauset2014} Virkar, Y., \& Clauset, A. 2014, Ann. Appl. Stat., 8, 89--119.

\bibitem[Wang et al.(1989)]{1989Sci...245..712W} Wang, Y.-M., Nash, A.~G., \& Sheeley, N.~R., Jr.\ 1989, Science, 245, 712

\bibitem[Wang \& Sheeley(1991)]{1991ApJ...375..761W} Wang, Y.-M., \& Sheeley, N.~R., Jr.\ 1991, \apj, 375, 761

\bibitem[Wang et al.(1991)]{1991ApJ...383..431W} Wang, Y.-M., Sheeley, N.~R., Jr., \& Nash, A.~G.\ 1991, \apj, 383, 431

\bibitem[Wang(2004)]{2004SoPh..224...21W} Wang, Y.-M.\ 2004, \solphys, 224, 21

\bibitem[Wheatland(2000)]{2000ApJ...536L.109W} Wheatland, M.~S. 2000, \apjl, 536, L109

\bibitem[Wheatland(2003)]{2003SoPh..214..361W} Wheatland, M.~S. 2003, \solphys, 214, 361



\end{thebibliography}
\end{document}